# Initial evaluation of miniature ultra-high-field commercial stellarator reactors with breeding external to resistive coils

V. Queral[1*], E. Rincón[1], A. de Castro[1], I. Fernández-Berceruelo[1], I. Palermo[1], A. Moroño[1], V. Tribaldos[2], J. M Reynolds[2], D. Spong[3], S. Cabrera[1], J. Varela[4]

[1] *Laboratorio Nacional de Fusión, CIEMAT, 28040 Madrid, Spain.*

[2] *Departamento de Física, Universidad Carlos III de Madrid, Leganés, 28911 Madrid, Spain.*

[3] *Oak Ridge National Laboratory (ORNL), Oak Ridge, Tennessee 37831, USA.*

[4] *University of Texas, Austin, Texas 78712, USA.*

* Corresponding author: V. Queral
vicentemanuel.queral@ciemat.es

**Abstract**

The working parameters and challenges of ultra-high-field pulsed commercial stellarator reactors of small plasma volume with breeding external to resistive coils (*transposed* stellarator) are studied. They may allow production of commercial heat and electricity in a tiny and simple device, and contribute to the knowledge on burning plasmas.

The concept is based on the previous works (V. Queral et al.) performed for the high-field experimental fusion reactor i-ASTER (J. Fus. Energy 37 2018) and the recent Distributed Divertor concept (non-resonant divertor on the full toroid; J. Fus. Energy 44 2025). The present proposal is driven by the limitation on the minimum size of typical commercial stellarator reactors due to the space needed for internal breeding and shielding of superconducting coils. This limit is about 400 m³ plasma volume, as deduced from e.g. ARIES-CS, ASTER-CP-(IEEE Trans. Plasma Sci. 52 2024) and Stellaris reactors. This fact, together with the accuracy and complexity of the systems, hinders quick iterations for the fast development of stellarator reactors.

The concept is based on a pulsed high-beta large-aspect-ratio stellarator of small plasma volume (2−4 m³) and ultra-high magnetic field (~ 10−20 T), composed by an external monolithic coil support and internal resistive coils (alike i-ASTER and UST_3 stellarators) of high neutron transparency, thermally-adiabatic conductors, a low-recycling Distributed Divertor to extract the huge short-pulsed heat power from ionized particles (pulse length ~ 5 $\tau_E$), low pulsed duty cycle of 1−5%, and liquid breeding material around and externally to the reactor core. Different cases and operating points are studied. The main elements, e.g. heat power on the Distributed Divertor, mechanical stresses in the coil support, radiation lifetime, and the prospect of net electricity production are evaluated. The involved challenges, impacting the potential feasibility of the concept, are assessed.

**Keywords**: stellarator, commercial reactor, miniature, ultra-high-field, external breeding, distributed divertor, resistive magnets.

## 1. Introduction

Magnetic-confinement fusion reactors may be classified in experimental and commercial fusion reactors. The first are usually called burning plasma devices and their main objective is the scientific research, i.e. ITER. The latter seek commercial profit and would have to compete with other electricity sources, like renewables or fission, or compete with the production of other services or products, e.g. heat or hydrogen production.

Stellarators are an important alternative to tokamaks for energy production due to their intrinsic steady-state, the absence of disruptions and fewer interlinked coils and systems. Nevertheless, stellarators are still less developed than tokamaks.

Among the diversity of matters studied in fusion research during decades, the reduction of size of commercial fusion reactor designs has been thoroughly pursued, i.e. Ref. [Gue 08] cites the progress performed (for relatively old designs).

For stellarators, the size of ARIES-CS was optimized and reduced to the minimum by reducing the breeding space, balancing the thickness of breeding+shielding in each sector, and considering an improved neutron shielding (WC) at certain regions of the inboard curved sectors [Gue 08]. The plasma volume for the reference design of ARIES-CS is 444 m³ [Lyo 07]. Recently, the high-field reactor ASTER-CP, based on centrifuge liquid walls (500 m³) [Que 24], the Stellaris reactor (425 m³) [Lio 25] and the compact fusion blanket stellarator reactor (smaller plasma volume) [Pro 24] have been defined. Certainly, the thickness of the breeding materials and of the radiation shielding are the main limitations to the size reduction in stellarators. For standard tokamaks, the space for the coil casings at the inboard is an additional limitation to the size reduction, due to: the typical lower aspect ratio of tokamaks, the existence of the central solenoid and the typical geometrical arrangement of the coils [Fed 24]. Recent advances in neutron shielding [Seg 22] have only modest impact on the thickness of the neutron shielding, due to the approximate exponential law of neutron and gamma penetration. This fact hinders further size reduction in stellarators.

The present work focuses on the reactor being commercially viable, that is, for profit. Nevertheless, the possible scientific progress obtained from this reactor would be also valuable. In particular, the work is dedicated to reactors of stellarator type.

In view of the difficulty for considerable size reduction of superconducting reactors with internal breeding, the present work studies the working parameters and challenges of ultra-high-field miniature commercial stellarators having the breeding material external to resistive coils. This concept is called ***μStellarator*** (***microStellarator***). These resistive stellarators will be commercially feasible if they reach profitability in some market niches, e.g. heat production, or $^3$He/tritium market. Miniature and simple resistive commercial reactors would also be decisive for quick iterations of successive designs and for enhanced physics knowledge, both essential for the fast development of stellarator commercial reactors up to maturity.

Indeed, the rapid scaling-up of construction of commercial nuclear reactors is much simpler and economic if small sizes and simple commercial designs are developed. For example, the first commercial civil (electricity production) fission reactors (1950's) had 5 MWe and ~100 m$^3$ of reactor vessel for the APS-1 Obninsk [IAE 26]) and 60 MW$_e$ ~70 m$^3$ for the Shippingport Atomic Power Station [Spr 23]. Improving and learning from quick iterations was rather straightforward in fission energy. Thus, construction of commercial fusion reactors, while using a reasonable investment, is an important factor in fusion energy development, given the limited available funds for the successive iterations. Miniature commercial stellarator reactors with breeding external to resistive coils (μStellarators) could contribute to ameliorate this difficulty, accelerating the innovation cycle.

Hypothetically, the concept might be also applicable to standard or spherical tokamaks under certain specific arrangements and specifications.

Certainly, the old Riggatron™ tokamak concept [Ros 84] has similarities to the present proposal, like the breeding material located outside resistive coils, very high magnetic field, the extremely harsh environment in the reactor vessel, and the need of frequent replacement of the reactor core. However, one important difference is the relatively long pulse ($\sim$ 120 s) in Riggatron. This fact, in Riggatron, may: **a)** hinder net energy production due to extreme cooling parameters in thermally-steady-state regime (i.e. at divertor targets), **b)** result in low reliability and availability in a high radiation environment (i.e. due to reliability of cooling microchannels) and, **c)** imply high cost of the (re)fabrication of a complex reactor core due to fast cooling (i.e. microchannels). Also, Riggatron would retain the typical drawbacks of tokamaks, such as disruptions and the usual need of continuous current drive during the pulse, which are enhanced by the extreme high power density of this device. These issues do not happen in μStellarators due to the short pulse and adiabatic heating of the components (thus, simple off-pulse cooling), and the absence of plasma current, disruptions and current drive in μStellarators.

A previous experimental device (built in the reality), somewhat resembling μStellarators, is the Scylla-closed theta-pinch [Sch 24][Can 74]. Though it is not fully comparable since it was not a stellarator, it achieved $\sim$ 20 T in a large aspect ratio device. Indeed, plasma confinement, MHD stability and plasma-wall interaction are different in stellarators.

The μStellarator concept is based also on the recent discovery (not yet proven experimentally) of the Distributed Divertor and Equi-power surface [Que 25]. Nevertheless, island divertors with detachment, high radiation fraction and sweeping may also be suitable.

An important antecedent for the current work is the concept of high-field ignition experimental fusion reactor and the resultant i-ASTER concept [Que 18]. Indeed, the average magnetic field on the magnetic axis in i-ASTER is $B \sim 10$ T for aspect ratio $A = 6$. However, the i-ASTER design is a burning plasma device, not a commercial fusion reactor.

An initial exploration of the working parameters and the difficulties of the main elements of μStellarator reactors is performed in the present work. It does not pretend a detailed physics and engineering design. It mainly intends the definition of the concept, and the evaluation of the challenges involved, which impact on the potential feasibility of the concept.

The article is organized as follows: Section 2 outlines the justification and methodology, Section 3 presents a geometrical sketch of the concept, Section 4 shows plots indicating potential operating points for the concept, Section 5 deals with the power load on divertorial surfaces, the power dissipated in the resistive magnets is estimated in Section 6, the stress in the monolithic coil support structure is studied in Section 7, neutronics matters, like lifetime and activation are estimated in Section 8, Section 9 discusses the challenges, uncertainties and further studies required, and finally, Section 10 mentions the advantages of μStellarators.

## 2. Justification, Methods and Assumptions

### 2.A. Justification for the proposal of a transposed pulsed resistive miniature stellarator

The difficulty to reduce the size of a superconducting commercial stellarator reactor to significantly lower than $\sim$ 400 m$^3$ has diverse cost implications for a fusion plant. Certainly, it implies difficult cost reduction of the containment buildings (i.e. see [Bro 18]), of the large remote maintenance systems, the hot cells to store the replaced blankets [Bro 18], and the cost of the reactor core itself. Also, the large reactor size hinders a fast iteration of successive reactors with reasonable investment, since they are accurate and complex nuclear devices, thus, expensive. Consequently, the present concept is envisaged to reduce some of such issues.

The essential drawback of resistive magnets in stellarator (or tokamak) power plants is the consumption of power in the coils. Indeed, part of the generated electricity does not reach the grid. Thus, larger heat exchangers, and larger turbines and generators are required in comparison to superconducting magnet devices, which impact on cost. The fractional power lost in the coils, and thus, the fractional recirculated power, depends on beta limit, enhancement factor of the energy confinement time $h_E$, plasma volume, aspect ratio and the quantity and type of conductors, as shown in this work. Advantageously, turbines and electric generators are low activated components, which have less impact on cost.

Also, the fractional recirculated power depends on the power consumed to pump the cooling fluids for divertors and first wall, which depends on the size and fusion power of the device. μStellarators hardly can work in continuous regime if net energy production is required. Otherwise, the cooling pumping power and the heat power density to be extracted in continuous regime at the divertorial surface and the coils could be excessive for net energy production. Pulsing is a drawback in some ways, but, the possibility of a feasible miniature fusion reactor producing abundant net energy could compensate the drawbacks of a pulsed regime.

Certainly, stellarator commercial reactors equipped with resistive coils and working in permanent regime appears feasible for large plasma volumes. For example, it is deduced from Ref. [Que 18] if the major radius is increased 2-fold (plasma volume $V_p \sim 240$ m$^3$) to allocate space for the breeding material. Also, for standard tokamaks, the feasibility may be deduced from Ref. [Woo 98], that shows no constraint on (low) beta limit for net energy production, and demonstrates that recirculated power and power for cooling the Joule-heat effect in the coils are not an issue. But, definitely, this is a huge resistive tokamak. For spherical tokamaks, a resistive reactor was proposed as one alternative of pilot plant [Men 11].

In μStellarators, the plasma volume cannot be large (i.e. < ~10 m$^3$) to have a thin *shell* for enough neutron transparency for tritium production. The term '*shell*' is defined as the wall of the toroidal coil support plus the layer of coils. Also, the fusion power density is high due to the ignition condition (Q = ∞) and the net electricity production requirement. Thus, the power exhaust dictates an operational scenario for the Plasma Facing Component (PFC) surfaces with very high loads, even for the case of a Distributed Divertor.

Actually, resistive coils have the next **advantages**:

- High positioning accuracy of resistive magnets is easier and cheaper to achieve than with superconducting coils. Indeed, superconducting coils experience large movements during cooldown and require delicate and accurate cryo-isolated legs and mechanical systems.
- Resistive magnets allow simpler and faster remote maintenance operations. For example: warm-up/cooldown of superconducting coils, which lasts weeks/months to decrease internal stresses, is avoided, which speeds up some processes; also, splitting the stellarator sectors [Que 18] is simpler for resistive coils due to the absence of thermal shields and the cryostat [Pal 26].
- Resistive magnets do not quench, decreasing the risk of coil failure.
- Activated resistive coils contain almost pure materials, which are simpler and cheaper to recycle and reuse than superconducting coils. Superconducting materials contain metallic alloys or ceramics, which are more difficult to recycle.
- Reduction of initial capital cost for the coils and simplicity of the power plant. This is essential for the first commercial power plants.

**2.B. Methods and assumptions. Terminology.**

The present work has two main objectives: **i**) study the working parameters of transposed miniature stellarator reactors under ignition condition for a wide range of plasma volumes, **ii**) analyse the difficulties, advantages and potential feasibility of the design.

The equations and methods established in Ref. [Que 18] are utilised to calculate and estimate the reactor physics and engineering parameters. However, presently, the equations are applied to various aspect ratios and to smaller plasma volumes. The aspect ratio is $A = 6$ for i-ASTER [Que 18]. As cited, i-ASTER is a pulsed experimental reactor under plasma ignition, which is not aimed at commercial energy production.

Electric effects due to pulsed regime and power consumption during pulse ramp-up/down are out of scope of the present study.

The method for the development of the work is composed of three elements:

**First**, a relation between the average beta limit $<\beta>_{\mathrm{lim}} \equiv \beta_{\mathrm{lim}}$ and the aspect ratio $A$ is defined and taken. For this, several stellarators in the bibliography have been studied: QI (quasi-isodynamic) magnetic configurations of 2, 3, 6, and 9 periods, LHD, and NCSX. The best (higher $\beta_{\mathrm{lim}}$ for fixed $A$) QI configurations fulfil $A \approx 4/3\ \beta_{\mathrm{lim}}(\%)$, i.e. $\beta_{\mathrm{lim}} \approx 8.5\%$ for $A \approx 12$ in QIP6 (Quasi-Isodynamic with poloidally closed contours of constant $B$ of 6 periods) [Mik 04][Bei 11]. LHD achieved $A \approx \beta_{\mathrm{lim}}(\%)$ and NCSX might achieve even higher. Also, a QI magnetic configuration with aspect ratio $A \sim 30$ and $\beta_{\mathrm{lim}} \sim 20\%$ was theoretically obtained [Bov 08]. QI magnetic configurations are preferred for μStellarators due to simpler plasma positioning, particularly in a short-pulse stellarator. Certain calculations for NCSX ($A$ = 4.5) resulted in $\beta_{\text{lim}} \sim 7\%$ [Sug 04][Zar 07], which would imply $\beta_{\text{lim}}(\%) > A$. Second stability regimes of high beta 7–20% in compact stellarators were theoretically predicted [War 04][War 02], but, they are yet to be experimentally proven. Additionally, a large population of alpha particles may help MHD stability, as happened in LHD, and a high plasma density should help to stabilise Alfven modes. Thus,

$$\beta_{\text{lim}}(\%) = A$$

is considered as a reasonable middle point for the called '***Reference Cases'***. Two other relations (other Cases) are also studied, Section 4.A.

**Second**, the ratio γ between the electrical power generated by the device ($P_e$) and the electrical power consumed in the coils by Joule-effect heating ($P_{\text{coils}}$) is set

$$\gamma = P_e \;/\; P_{\text{coils}}$$

where $P_e = P_{\text{fusion}} / 3$ can be reasonably assumed. $P_{\text{fusion}}$ is the total power generated by fusion reactions, $P_{\text{fusion}} \equiv P_f$, Section 4.

γ is similar to $Q_{\text{eng}}$, (Engineering Gain Factor), but, γ includes electricity production (not net heat production) and does not include other pulsed energy consumptions. E.g. ECRH consumption to start-up the plasma up to ignition, small continuous pumping power for reactor-core cooling and electrical power corresponding to the Joule-effect during ramp-up of coils are not included in the γ definition.

For γ = 1 all the electricity generated would be consumed in the coils, resulting in zero net electricity production, but, still, abundant and valuable heat would be produced.

As a starting point for attractive enough electricity production

$$\gamma \approx 2$$

is considered. This means that 50% of the produced electricity is recirculated to feed the coils and 50% goes to the electricity grid.

**Third**, we set the confinement enhancement factor $h_E$ (as in the International Stellarator Scaling 2004, ISS04 [Wel 09]) as a linear combination of $\beta_{\text{lim}}$ (as unitarian, ∈ [0,1]),

$$h_E = \lambda \, \beta_{\text{lim}}$$

λ = 5 results in γ = 2 for $V_p$ = 4 m$^3$ and the conditions in the study. γ is about 15% lower for $V_p$ = 2 m$^3$ , but, this is still considered γ ≈ 2. $h_E$ is 1 and 1.4 for two relevant Reference Cases, Section 4.A. This values were achieved in W7-AS and LHD [Wel 09].

A parameter scan is performed for different $\beta_{\text{lim}}$, and so, for different aspect ratios and $h_E$. Most of the aspect ratios are relatively large. Large aspect ratio is favoured since the thickness of the layer of coils and monolithic coil support is thinner (higher neutron transparency) for the same plasma volume, and since $\beta_{\text{lim}}$ increases, which is important to decrease the power consumed in the coils. The range of parameter exploration is intentionally extensive to offer preliminary estimates of potential operating conditions for the design. Plots are produced for these wide range of parameters. The process is, to some extent, similar to the one followed in Ref. [Que 18].

### 2.C. Terminology

The term '*Transposed'* means that the location of the breeding material is external to the coils, contrary to the common internal location.

An *Equi-power surface* is a toroidal surface around the plasma, not conformal with the Last Closed Flux Surface (LCFS) and defined geometrically very precisely, that receives the same power density from ionized particles at any location. It was defined and one example calculated in Ref. [Que 25].

The term *μASTER_v1* is used for a particular μStellarator having certain specifications. In particular, its magnetic configuration has 8 periods.

*Neutron transparency* means the capability of the reactor core materials to transmit neutrons able to produce tritium at the exterior of the reactor core.

*Long term activation (LT)* is considered 1000 years after shutdown. In Sections 7.B, 8 and others, the nomenclature utilized for activation is, for example, Zr: ~20x LT Bq ; 100x LT γ. It means that the chemical element (Zr) has 20 and 100 times higher activation than Fe 1000 years after shutdown, and indicates the type of radiation (Bq ≡ Specific Activity ; γ ≡ Dose rate).

*Medium term activation (MT)* is considered 100 years after shutdown.

## 3. Scheme of the concept and dimensional parameters

Fig. 1 shows a scheme of the μStellarator, including: reactor vessel, external liquid breeding material, monolithic coil support, feeding pipes and auxiliaries. The reactor vessel is represented in transparent view to allow observation of the internal elements. The free surface of the breeding material is represented opaque while the liquid is represented transparent.

Fig. 2 shows a scheme of the monolithic coil support, the layer formed by coils/cables/conductors (*Coils-layer*), and the plasma. The cut is produced at the cross-section having bean plasma shape. The shell acts also as vacuum vessel.

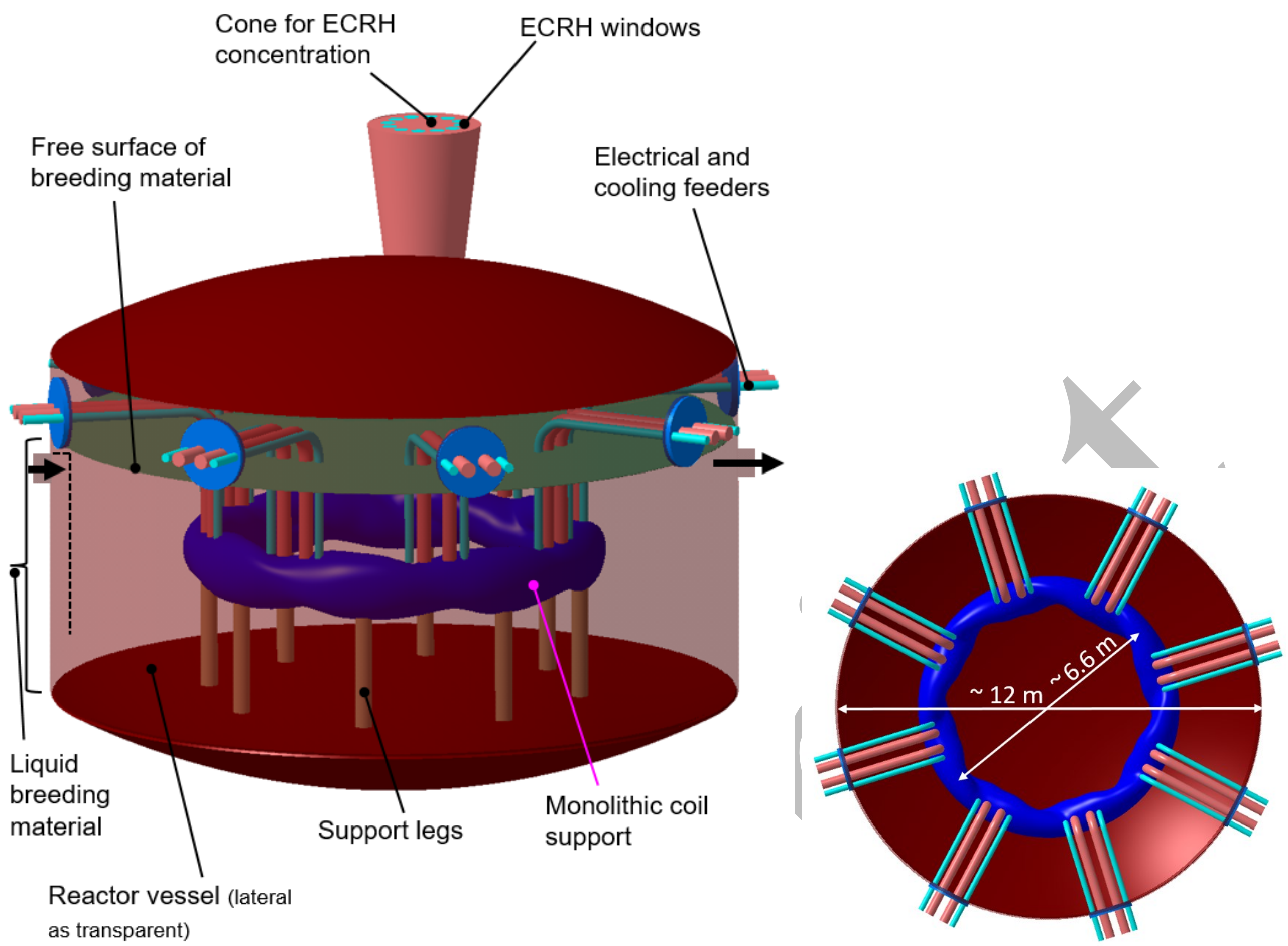

**Fig. 1** Scheme of the concept of a μStellarator commercial reactor. Perspective view (left) and a plan view (right, reduced size). The coil support representation comes from the magnetic configuration of a stellarator of $A = 20$ and 8 periods called μASTER_8_20_C43 (Section 3.B). The reactor head, the ECRH cone and the liquid breeding material is hidden in the plan view. The reactor vessel is filled with liquid breeding material.

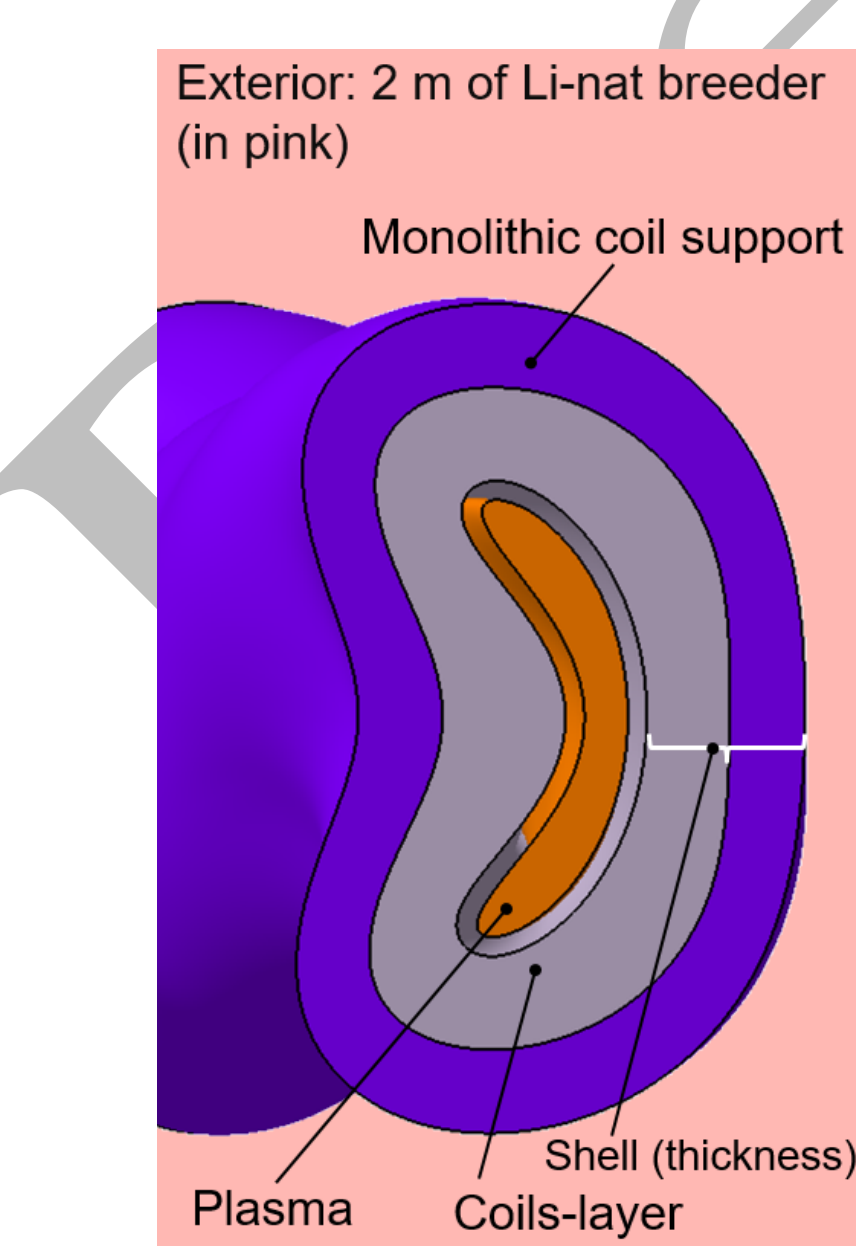

**Fig. 2** Scheme of a monolithic coil support (blue), the Coils-layer (grey), the plasma (orange) and the external liquid breeder (pink) at the bean shape of a magnetic configuration.

Fig. 3 depicts schematically the parameters involved in the design. $\xi$ is a factor relating the minor radius of the winding surface, $a_c$, to the plasma minor radius ($a_c = \xi\ a$). The distance from the interior of the Coils-layer to the plasma, $\Delta'$, is reduced as much as possible (proportionally smaller than in i-ASTER) to decrease the power consumed by the coils. Since $\Delta' = (\xi - \varepsilon/2 - 1)\ a$, $\xi = 1.6$ gives $\Delta' = 0.1\ a$, being $a$ the minor plasma radius.

'Relative thickness occupied by coils' $\varepsilon = 1$ and 'relative thickness of monolithic coil support' $\psi = 0.75$ are selected for the calculations in this paper. $\psi$ has to be thin in order to allow enough neutron transmission through the shell, and still give stresses compatible with the material.

The thickness of the shell of the reactor core is defined as $\delta = (\varepsilon + \psi)\ a$. In order to fix certain parameters of the design, after performing calculations on TBR in a spherical model, see Section 8.A, $\delta = 0.3$ m is taken. This is the maximum thickness of the shell that allows TBR ≈ 1 for certain materials and isotopes. Thus, a ≈ 0.17 m ($\delta$ / ($\varepsilon + \psi$)). For a = 0.17 m: $V_p$=2 m$^3$ gives $A \approx 20$ ; $V_p$=3 m$^3$ $A \approx 28$ ; $V_p$=4 m$^3$ $A \approx 40$.

The external surface of the monolithic coil support and the feeders would be coated with SiC, tungsten or other material compatible with the liquid breeding material in the pool. Solid breeder is a back-up option.

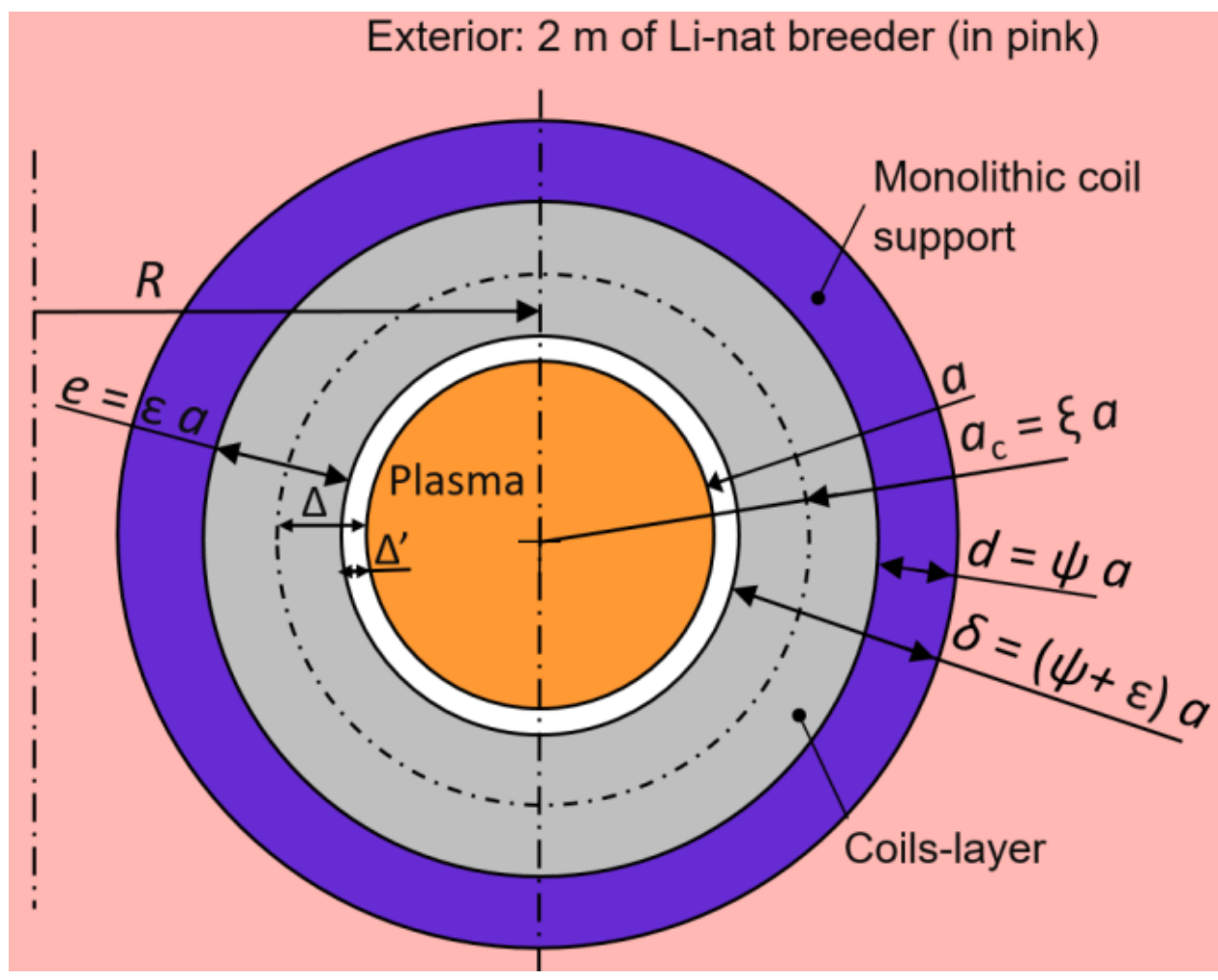

**Fig. 3** Schematic poloidal cross-section of the plasma, coils-layer, monolithic coil support and definition of the main dimensions utilised in the calculations. See Fig. 11 for Coils-layer details.

The (preferable) liquid breeding material will be stirred inside the reactor vessel to avoid excessive thermal gradients. The speed of the fluid at the inlet/outlet will be low since the section of the pipes can be as large as needed.

### 3.A. Material for the conductor

Aluminium and copper (particularly the isotope $^{65}Cu$) are the reference materials for the conductors, due to acceptable neutron transparency and high electrical conductivity. The calculations in Section 6 have been done for aluminium conductors with 5% fraction of insulation, and are also applicable for copper conductors where the *Inter-cable matrix* (Fig. 11) is about 40% of the Coils-layer volume, due to the higher electric conductivity of copper (Al has 0.65 IACS). The isotope $^{65}Cu$ is more neutron transparent than natural Cu and activates much less than Cu-nat, which are important advantages, Section 8.F. Copper is coherent with the Option '2' in Section 6. High electrical conductivity Aluminium alloys, being weaker than copper, additionally allow less space for the inter-cable matrix due to the lower conductivity. Aluminium is coherent with the Option '1' in Section 6. A comparison of the activation of Al and Cu with Fe is performed in Section 8.F.

Superconducting coils appear unfeasible for a commercial transposed miniature stellarator reactor. Other advanced materials like CNT (carbon nanotube) bundles or graphene-based coils also appear inconvenient.

The parameters taken for this initial study on µStellarators are summarised in Table 1.

| **Parameter** | |
|---|---|
| $a$, minor radius | 0.17 m |
| $\xi$, factor of coil radial position | 1.6 |
| $a_c$, minor radius of centre of the coil, $a_c = \xi\, a$ | 0.27 m |
| Δ', distance from interior of Coils-layer to plas. | 0.017 m |
| $\varepsilon$, relative thickness of the coils-layer | 1 |
| $\psi$, relative thickness of the coil support | 0.5–0.75 |
| $\delta$, thickness of the shell, $\delta = (\varepsilon + \psi)\, a$ | 0.255–0.3 m |

**Table 1** Parameters taken for this initial study on µStellarators, in particular corresponding to the reactor µASTER_v1. Some variations are possible for other µStellarators.

### 3.B. Preliminary magnetic configuration of high *A* and high beta

The main objective of this magnetic configuration is to prove that magnetic configurations of the type required for µStellarators exist.

The search of the magnetic configuration has been produced by the CASTELL code. CASTELL comprises a collection of Information Technology tools and codes that are integrated and linked together. This suite includes, among other components, a generator for guiding center orbits, routines for the geometric handling of toroidal surfaces, and an approximate calculator for neoclassical confinement. In addition, it is capable of automatically invoking external codes such as NESCOIL [Mer 87], the variational moments equilibrium code (VMEC) [Hir 83], and the code for ballooning rapid analysis (COBRA) [San 00]. In earlier works, CASTELL was employed to find magnetic configurations for the UST_1, UST_2, and UST_3 stellarators [Que 21][Que 15b], as well as for the ASTER-CP reactor [Que 24].

The configuration QIP6 [Sub 06] (QI of 6 periods), kindly supplied years ago (see acknowledgements), was utilized as initial configuration (*seed*) for the optimization. QIP6 has $A \approx 12$. Gradually the number of periods and the aspect radio was increased, while optimizing/increasing the MHD stability by using VMEC and the optimization method BOBYQA implemented in CASTELL. First, fixed boundary VMEC was utilized and, finally a second optimization process was produced in free-boundary, to save computing time. The level of neoclassical confinement was calculated at the end of the MHD optimization process by MOCA code [Tri 01] and is about 10 times lower than in QIP6. This is still too low, since it is slightly lower than the LHD inward configuration [Bei 11]. Indeed, neoclassical confinement of the level of LHD inward might be considered too low for a reactor [Bei 21]. Nevertheless, only reasonable neoclassical confinement is sought for µStellarators, since high beta limit is vital in this type of stellarator reactor. Thus, high beta limit is prioritised.

A magnetic configuration of aspect ratio $A = 20$ and Mercier and ballooning MHD stable up to $\beta_{\rm lim} \approx 14\%$ was found by CASTELL code as described, called µASTER_8_20_C43. This is calculated for a VMEC pressure profile of $p(s) = K\,(1\text{-}1.5s\text{+}0.5s^2)$. $\beta_{\rm lim}$ = 16 % corresponds to Case A''$_{20}$, see Section 4.A.

Fig. 4 shows the Poincaré plot in vacuum (in colours) obtained from coils and the input LCFS (in black squares). Coils are located slightly nearer (~0.07 m instead of 0.1 m) to the LCFS than $\Delta = \xi\, a - a$ , see Fig. 3. This distance should be slightly increased in the future.

Islands at the plasma edge should be small for the Distributed Divertor, as also sought in other non-resonant divertors [Bad 17], to avoid the islands crossing the Equi-power surface.

The rotational transform, $\iota$ , in free-boundary VMEC is in the interval [1.2-1.58] for beta 13%, and in the interval [1.45-1.56] for $\beta$ = 0%. In fixed boundary (from LCFS, not from a set of coils), the rotational transform is in the interval [1.47-1.62] in vacuum. Rotational transform below 1.6 was sought to avoid a large magnetic island appearing at the rational 8/5 = 1.6 (the configuration has 8 periods). The rational 8/5 could be useful if an island divertor were decided, though the reference divertor for μStellarators is a non-resonant 'Distributed Divertor' [Que 25]. The island from the rational 3/2 (in vacuum) is small or inexistent in the Poincaré plot, which is positive. This behaviour would require further study in vacuum and at high beta.

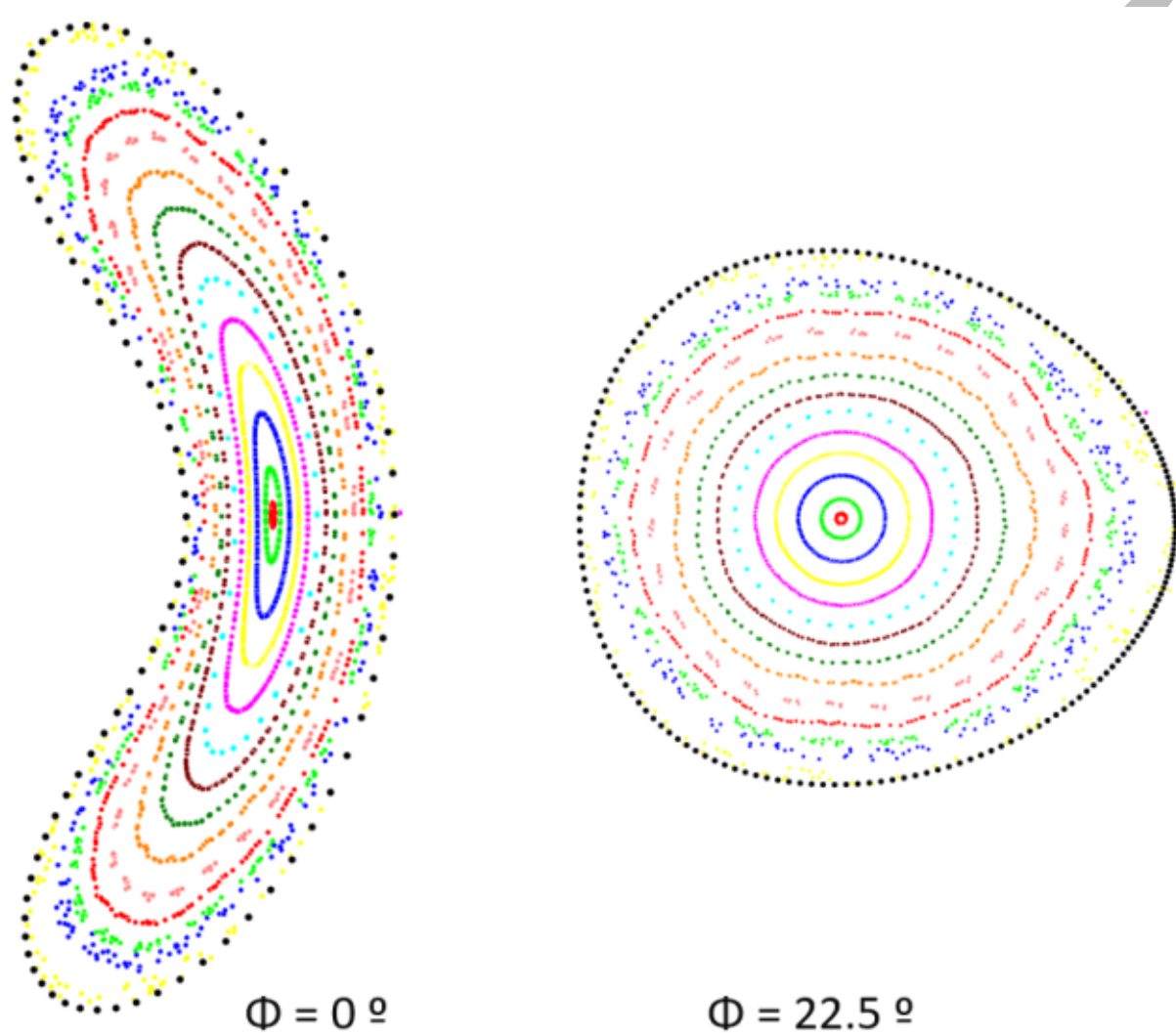

**Fig. 4** Poincaré plot in vacuum for the magnetic configuration μASTER_8_20_C43 of 8 periods. The same coils as for the free-boundary VMEC calculation are utilized. The original LCFS is represented in black. Note that the Poincaré plot comes from coils, not from the represented LCFS.

Thus, the main result is that at least one magnetic configuration exists of the type required for μStellarators. Also, it may serve as a starting point for further optimization.

An optimization process starting from a broad space of 'seed' configurations would be necessary to try to find a high $A$ and still higher $\beta_{\rm lim}$ configuration with reasonable neoclassical confinement.

## 4. Physics parameters and plots

Rotational transform $\iota$ = 1.4 is considered, which is an intermediate point in the interval of rotational transform in the magnetic configuration μASTER_8_20_C43, obtained in this work.

The plasma core dilution factor is considered $f_{\rm d}$ = 0.9. The existence of lithium on the Equi-power surface (all the internal surface of the device would act as a lithium divertorial surface [Que 25]), may allow such level of plasma purity, as justified in [Que 18]. Indeed, the lithium radial penetration to the core is expected to be partially damped by the fact of the fast Li ionization and potential of very high prompt redeposition of eroded lithium [Bro 01]. This low content of lithium in the core plasma has been observed in many experiments with lithium PFCs, both in stellarators and tokamaks around the world [Cas 21].

A multiplicative factor $k_\beta$ for $\beta_{\rm lim}$ is established, $k_\beta$ = $2^{1/2} \approx 1.414$, to generate the different possible Cases in the plots. The $n$ Case in the plot has $\beta_{\rm lim} = n\, k_\beta\, \beta_{\rm lim,1}$. And here we take the first value of $\beta_{\rm lim,1}$ = 10%. Five values of $\beta_{\rm lim}$, starting from $\beta_{\rm lim,1}$ = 10%, are considered. It represents a broad range of parameters. Consequently, the plots in the current and next sections include potential feasible and unfeasible Cases.

Also, $A = \beta_{\rm lim}(\%)$ and $h_{\rm E} = 5\, \beta_{\rm lim}$ (which gives $\gamma \approx 2$) is considered (see Section 2.B). For each combination of $\beta_{\rm lim}$, and the resultant $A$ and $h_{\rm E}$, and the other parameters taken, we estimate the minimum magnetic field required for ignition (Fig. 5), following the same procedure as in Ref. [Que 18].

### 4.A. Ignition magnetic field *B* for the different Cases

Fig. 5 shows the minimum magnetic field needed for ignition for each combination of $A$, $h_{\rm E}$ and $\beta_{\rm lim}$ established in Section 2.B and 4. Each combination is called '*Case*' and they are denoted with the labels $\mathbf{A_{10}}$ to $\mathbf{A_{40}}$. The sub-index indicates the aspect ratio of such particular μStellarator reactor.

$\lambda$ = 5 and $A = \beta_{\rm lim}(\%)$ generate the *Reference Cases*, labelled in black and larger font in Fig. 5.

The violet line crossing the different cases indicates devices with $a \approx 0.17$ m. Potential operating points (some are presently unfeasible) for this $a$ value for each Case are marked with a cross. If other $a$ value would like to be considered, the violet line would be translated to the left or to the right. For example, for $a$ = 0.14 m the line would cross the point of $V_{\rm p}$ = 0.57 m$^3$ for $A$ = 10, Case $A_{10}$. For $a$ = 0.17 m the relation of $V_{\rm p}$ and $A$ is indicated at the beginning of Section 3. Note that $\delta$ (thickness of the reactor shell) and $a$ are related by $\delta = (\varepsilon + \psi)\, a$ (Section 3).

Plasma physics in various types of magnetic configurations (QA, QI, QH) might be improved to hypothetically achieve $\beta_{\rm lim}(\%) > A$ at high $A$ values. Indeed, it was theoretically obtained in NCSX for low $A$ [Sug 04][Zar 07].

Certainly, the physics parameters can be (properly) varied while keeping most of the geometrical engineering design. The engineering parameters change according to the physics parameters, as shown in the plots in the next sections.

As points of reference, the plasma volume and magnetic field for larger devices with relatively long pulse is: in IGNITOR $V_p \sim 10$ m$^3$ and $B \sim 13$ T, in i-ASTER $V_p \sim 30$ m$^3$ $B \sim 10$ T.

In all the plots $V$ is indicated for plasma volume instead of $V_p$.

Undeniably, some of the values of magnetic field appearing in the Fig. 5 are not viable with the current technical means, e.g. tensile strength of the best materials or heat load on the divertorial surface (see next sections). However, some of the currently impossible parameters might be feasible in the future, due to improvements in materials and plasma physics.

In relation to that future possibilities,

***Optimized-Cases*** ≡ ***Optim-Cases*** are defined as cases under the assumption that $\beta_{lim}$(%) could be higher than the aspect ratio $A$ by a factor 1.26 (1.26 $A$ = $\beta_{lim}$(%). They are indicated in green and single quotes in Fig. 5 and have $\lambda \approx 5.2$ ($\lambda = 5.195$). Note that from Section 2.B $h_E = \lambda\, \beta_{lim}$. These particular relations also give $\gamma \approx 2$, which is the objective of the μStellarator. The physics parameters of the optim-Cases are difficult to obtain while the engineering ones are more straightforward.

***Pre-Cases*** are preliminary cases having $A = 1.26$ $\beta_{lim}$(%). This relation was almost achieved for the QI configuration μASTER_8_20_C43 obtained in this work, see Section 3.B. One pre-Case is indicated in Fig. 5 in red and double quotes. They have $\lambda = 4.81$. These preliminary cases should be improved in the future to achieve at least the Reference Cases. The physics parameters of the pre-Cases are relatively simple to obtain theoretically while the engineering parameters are challenging for most of the pre-Cases.

The optim-Cases and pre-Cases, which are not indicated in the plot, can be deduced from the cited relations.

The average magnetic field at the magnetic axis for three cases is: $B$ = 10 T for Case $A_{40}$, $B$ = 13.5 T for Case $A_{28}$, and $B$ = 18.8 T for Case $A_{20}$.

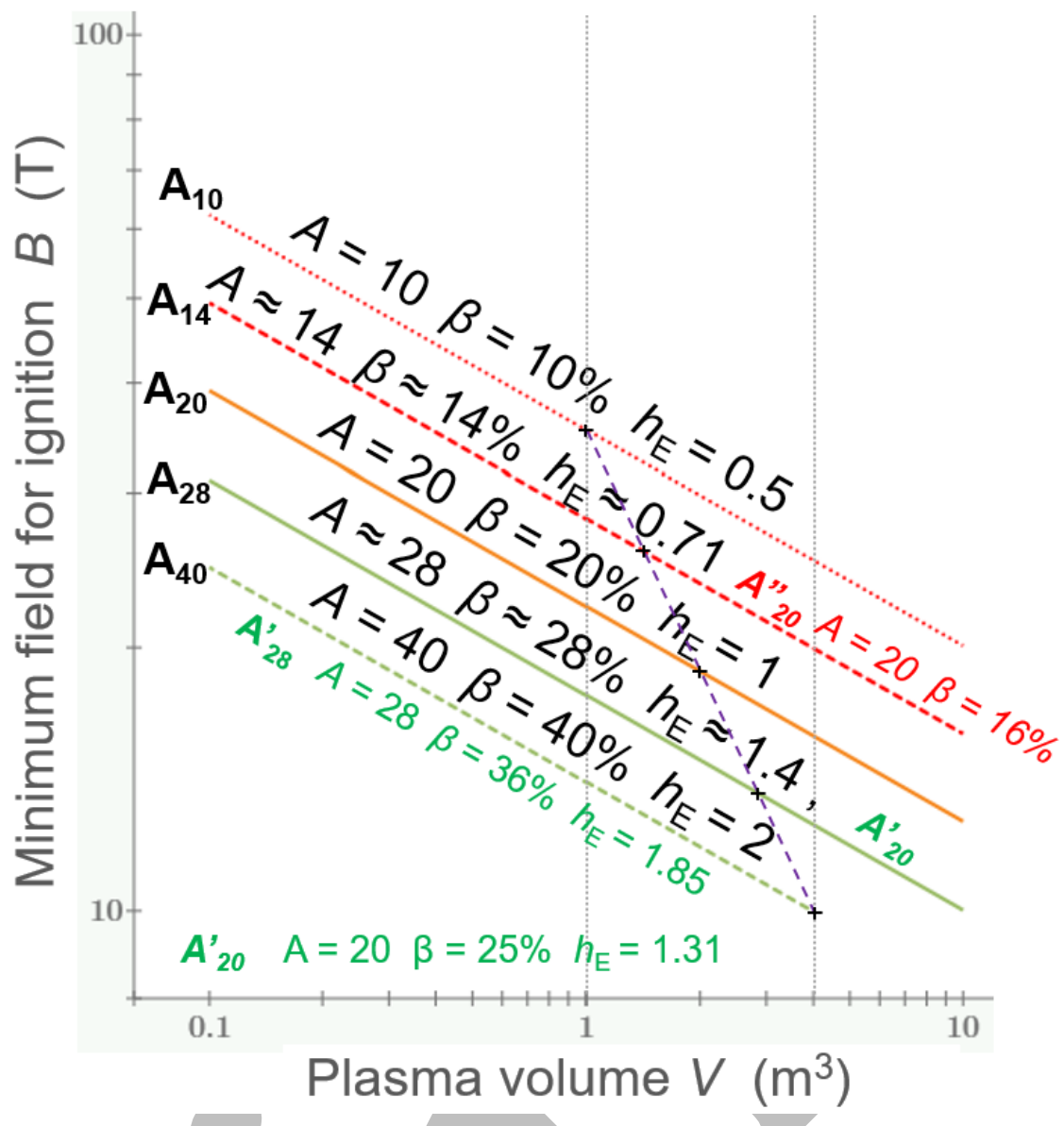

**Fig. 5** Minimum magnetic field $B_0$ for ignition of the plasma in μStellerators for the reference set of parameters (*Reference Cases*) defined in Section 2.B and 4 as a function of the plasma volume. $\gamma \approx 2 \equiv 50\%$ net electricity production. Different curves correspond to different assumptions on beta limit $\beta_{lim}$, which defines $A$ and $h_E$. The violet line crossing the different Cases (the red/orange/green lines) indicates devices with $a \approx$ 0.17 m ($a$ = 0.1717 m) ($\delta \approx 0.3$ m for $\psi = 0.75$ and $\varepsilon = 1$). Optim-Cases and pre-Cases are indicated in green and red fonts and have $A \neq \beta_{lim}$(%). Only three of such optim/pre-Cases are indicated to avoid muddling the plot. Case $A''_{20}$ has $h_E = 0.76$.

### 4.B. Density and temperature needed for ignition, fusion power, pulse length

The line density $n_{line}$ is ≈ 1, 1.2 and 1.7 × $10^{22}$ m$^{-3}$ for the Cases $A_{40}$, $A_{28}$ and $A_{20}$ respectively. The line-averaged density needed for ignition results lower than the Sudo limit. The results are obtained following the method in Ref. [Que 18]. The line density is about one order of magnitude higher than the maximum density achieved in LHD (peak density 1.2 × $10^{21}$ m$^{-3}$ at 2.5 T [Yam 11]). Such relatively high density in μStellarators might be achievable at fields higher than 13 T.

The pulse length $t_p$ is selected $t_p = 5\,\tau_E$. $t_p$ has to be balanced between the need of a large pulse to minimize the fraction of Joule-effect ramp-up/down losses and the ECRH energy losses (up to reaching ignition), and the need to be short to avoid overheating the conductors/insulation of the cables and the Equi-power surface. The ECRH energy used to reach ignition has to be minimized following the proper heating path (Cordey-pass), similarly to proposed for the HELIAS reactor, i.e. [War 15]. The cut-off density for 2$^{nd}$ harmonic O2 ECRH heating is 7 × $10^{21}$ m $^{3}$ for 13.5 T (Case $A_{28}$, at 750 GHz), somewhat lower than the line density. O3 ECRH for the last heating up to ignition might be necessary. Gyrotrons at such high frequencies are experimental and still have low efficiency, but high short-pulsed power can be achieved. Different ECRH heating modes may be needed along the Cordey-pass.

As an initial guess $t_p = 5\,\tau_E$ may give a satisfactory balance. The optimization of how many energy confinement times should last the pulse for maximum energy production and lower recirculated energy is out of scope of the present work. The relation $t_p = 5\ \tau_E$ is the same as the one utilized in i-ASTER [Que 18].

In particular, $\tau_E = 0.042$ s for Case $A_{40}$ ($t_p = 0.21$ s), $\tau_E = 0.033$ s for Case $A_{28}$ ($t_p = 0.16$ s), and $\tau_E = 0.024$ s for Case $A_{20}$ ($t_p = 0.12$ s), obtained from the scaling law ISS04 [Wel 09].

Fig. 6 indicates the fusion power generated $P_f$ ($P_f \approx 5\ P_\alpha$). Conservatively, the power produced in the breeding material by nuclear reactions is not included, that is, the energy multiplication factor in the breeding material is considered 1. The fusion power is nearly constant with respect to the plasma volume for the studied cases. This is mainly caused by the reduction of the minimum $B$ for ignition at larger plasma volume (Fig. 5) and the equations involved.

The ignition temperature is independent of $A$, $\beta_{lim}$, $h_E$, and $V$. For the assumed $Z_{eff}$ and pressure profile (it is comes from the temperature and density profile for HELIAS reactor plotted in Ref. [Wob 03]), the central plasma temperature is $T_{0.ig} = 13.7$ keV.

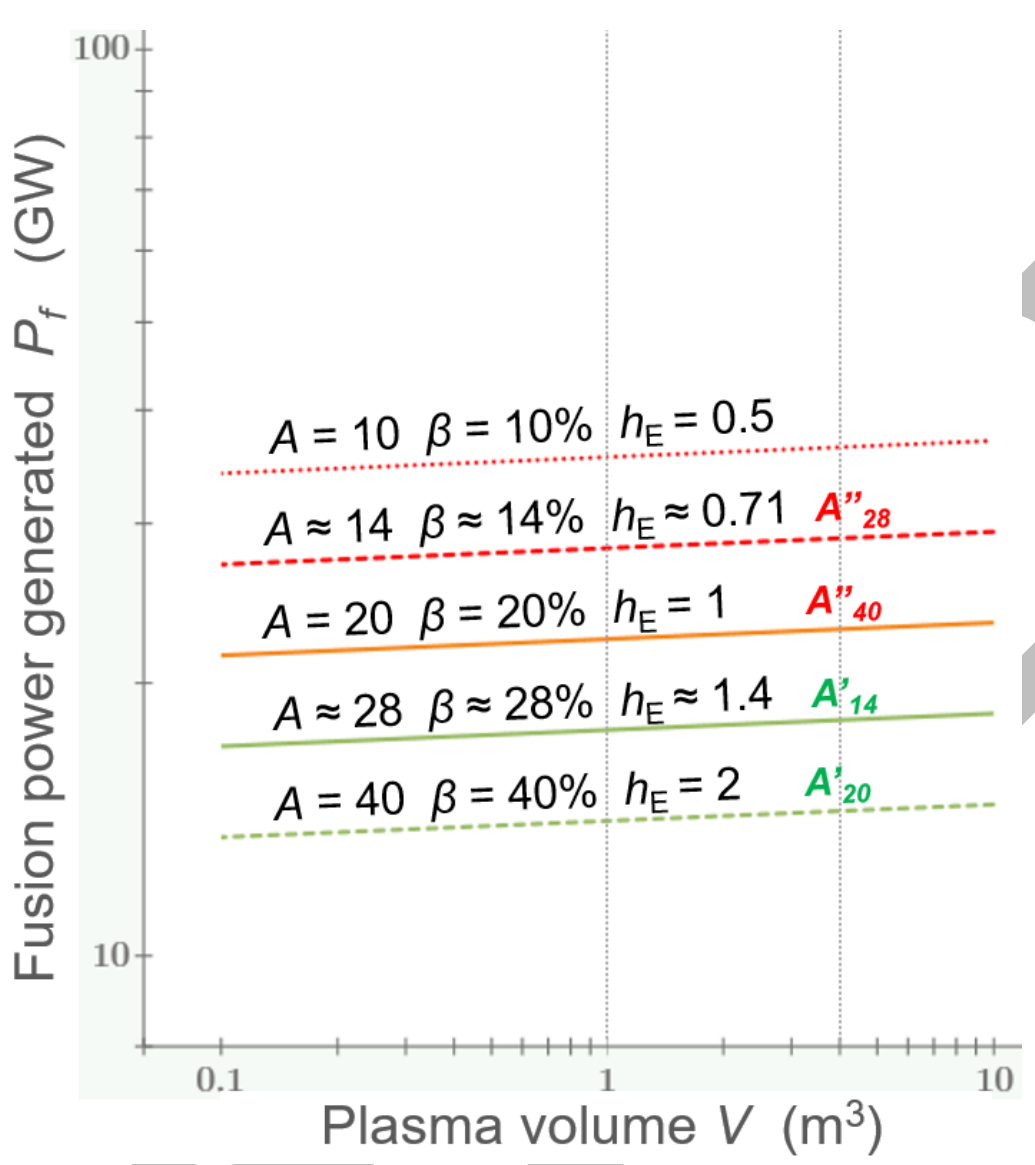

**Fig. 6** Fusion power generated $P_f$ for the combinations of $A$, $\beta_{lim}$, $h_E$ and minimum field for ignition presented in Fig. 5. Two pre-Cases and two optim-Cases are indicated.

The fusion power generated during the pulse is 22.6 GW for Case $A_{20}$, 18.1 GW for Case $A_{28}$ and 14.5 for case $A_{40}$. These values can be also observed in Fig. 6 as an approximation.

For the fusion power generated, the optim-Cases jump two lines downwards, and the pre-Cases two lines upwards. Thus, the fusion power generated can be also estimated from the plot for other non-Reference Cases. Only four optim/pre-Cases are indicated in Fig. 6 to understand the variation in fusion power generated for the optim/pre-Cases.

This result (jump of two lines) might be foreseen from the variation of the fusion power with $\beta^2$. It implies high sensitivity of this type of reactor to the beta achievable in relation to the aspect ratio.

The results fulfil the scaling law of the energy confinement time ISS04 [Wel 09]. This scaling law includes the turbulent and neoclassical transport components in the considered stellarators. The central plasma pressure in the $A_{28}$ μStellarator would be ~ 45 MPa. Comparatively, the tokamak Alcator C-Mod experimentally achieved 0.45 MPa at 5.7 T and similar minor radius [Hug 18]. Nevertheless, Case $A_{28}$ is defined with a magnetic field 2.4 times higher (13.5 T) and beta an order of magnitude higher than Alcator C-Mod, which may allow such high pressure gradient.

## 5. Power exhaust

The heat load per unit of surface on the divertorial surface (an Equi-power surface [Que 25]) $P_d$ is calculated by dividing the total incident power by the wetted area. The wetted area is related with the plasma surface $S_p$ by a 'concentration factor' $K_d$. The heating power $P_h$ received by the divertorial surface is a fraction or equal to the alpha heating power $P_\alpha$.

$$P_d = K_d\ \frac{P_h}{S_p}$$

$K_d$ depends on the particular magnetic configuration and divertor type. Indeed, in certain regimes $K_d = 18$ in LHD with an Intrinsic Helical Divertor and $K_d = 55$ in W7-X with an Island Divertor, [Que 18]. Differently, in μStellarator reactors the new type of non-resonant divertor, called Distributed Divertor [Que 25], is expected to provide much lower $K_d$.

We perform four **assumptions**: **1.** The full vacuum vessel surface acts as an Equi-power surface in a Distributed Divertor. The small opening for ECRH plasma start-up (Fig. 1) is neglected. The surface of the Equi-power surface is taken equal to the plasma LCFS ($S_p$) and is: for the Case $A_{40}$ $S_p$=47 m², $A_{28}$ 35 m², $A_{20}$ 24 m² (the average of the three is 35 m²) ; **2.** The possible non-uniformity in particle reception due to changes in the plasma and the Equi-power surface position are neglected (thus, the power from ionized particles is assumed uniformly distributed on the Equi-power surface) ; **3.** Bremsstrahlung radiation (i.e. 600 MW for Case $A_{28}$) is uniformly distributed on the Equi-power surface. **4.** Longitudinal ridges (shaped like the letter 'V', VVVV) in the direction of the magnetic field lines on the Equi-power Surface can be built. Either additive manufacturing or laser cutting could be possible manufacturing methods. It is assumed that this ridges are able to increase the effective surface twice.

A fraction of the power from ionized particles may be converted into line radiation. But, since the full vacuum vessel is a divertorial surface, the power received by the divertorial surface is still the total power. Thus, $P_h = P_\alpha$. Nevertheless, radiation is positive to reduce erosion of solid divertor targets and to decrease sputtering in liquid targets/surfaces.

To have an insight of the values, for the **Case $A_{28}$** ($a$ = 0.17 m, $S_p$ = 35 m$^2$), $P_\alpha$ ~ 3.4 GW (of which 0.6 GW results in Bremsstrahlung radiation) would be received uniformly by the Equi-power surface, resulting in $P_d$ = $P_\alpha/S_p$ ~100 MW/m$^2$ if there are not longitudinal ridges (that is, $K_d$ = 1), during 0.16 s .

For the **Case $A_{20}$** ($a$ = 0.17 m, $S_p$ = 24 m$^2$), $P_\alpha$ ~ 4.5 GW will be received approximately uniformly by the Equi-power surface, resulting in $P_d$ ~ 190 MW/m$^2$ if no ridges, during 0.12 s.

The **Case $A_{40}$** gives $P_d$ ~ 60 MW/m$^2$ following the same calculation, and $t_p$ = 0.21 s.

The resulting heat load on the divertorial surface under the assumptions $K_d$ = 1 and $P_h$ = $P_\alpha$ is plotted in Fig. 7.

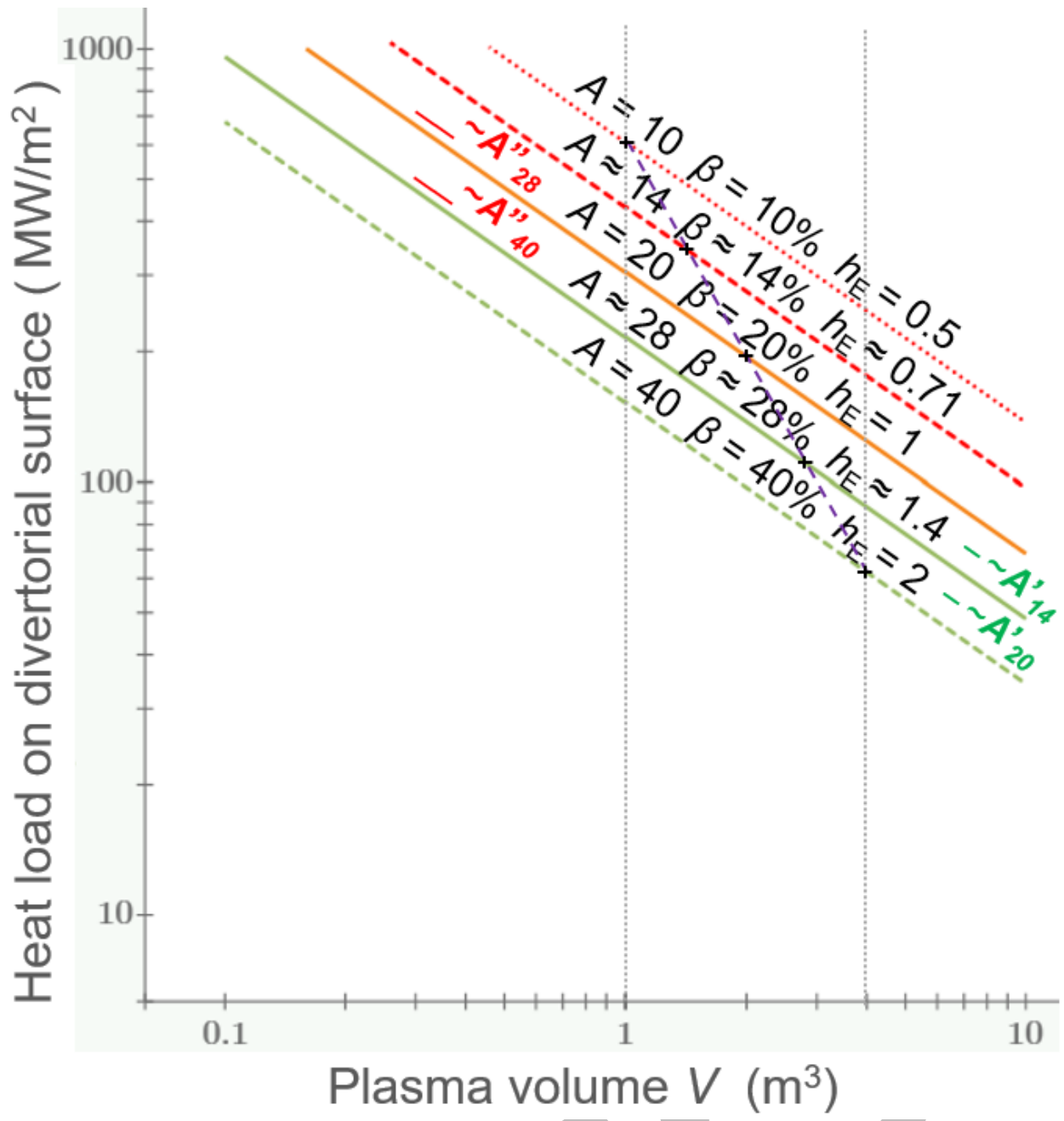

**Fig. 7** Pulsed heat load (ionized particles + radiation), $P_d$ , on the divertorial Equi-power surface of the Distributed Divertor for different Cases, for $K_d$ = 1 (≡ no ridges) and the full $P_\alpha$ flowing to the Equi-power Surface ($P_h$ = $P_\alpha$), for the minimum field for ignition presented in Fig. 5, plotted as function of the plasma volume and the combinations of $A$, $\beta_{lim}$, $h_E$. Some pre/optim-Cases are indicated.

Under these conditions, the pulsed (space-averaged) **neutron wall load** is four times higher than the heat power load on the divertorial surface shown in Fig. 7.

A thin (microns) film of liquid metal on a highly thermal conducting material (called substrate, Section 5.B) is the reference method for power exhaust in µStellarators . The layer of liquid has to be thin to avoid the relatively low thermal conductivity (no turbulence assumed) of the possible liquid metals compared with Cu, W and Al. This liquid metal layer would act as a protective, sacrifice surface, precluding the damage to the underlying substrate. Liquid films with high reactivity (this produce pumping and low recycling) and with low reactivity are studied in Section 5.A.

Refilling-rewetting or re-coating the PFC each certain number of pulses would be necessary. Such refilling has been proven in the case of repeated testing of tin liquid metal PFC exposed to pulsed high-heat fluxes relevant for reactors, like the one pursued herein [Cas 25]. Certainly, the pulsed regime simplifies the matter.

**Flowing liquid lithium** is considered a back-up method for power exhaust. Lithium jets, high speed lithium droplets, film flows or divertorlets [Sae 22] for reactors were studied in Ref. [Que 25]. Following such calculations, for $\Delta T_{surf}$ = 200 ºC, the exposure time for lithium should be ~ 3 x 10$^{-4}$ s, which is rather short. Only special divertorlets or other millimetre-size exposure-length methods would be possible. Thus, for µStellarators, flowing liquid metals are considered less appropriate than static materials.

### 5.A. Materials for the film located on the Equi-power substrate

Liquid and solid materials potentially appropriate as thin film, with and without pumping effect, are described next.

● **Liquid films for pumping and low recycling**

A thin film (microns) of liquid lilthium on a highly thermal conducting material (Cu, W or Al, called substrate) could be conceived. The liquid lithium could be located in W nets/mats (most common Capillary Porous System (CPS) [Evt 99]), porous W or other capillary materials. W does not need a protective coating to avoid corrosion of the substrate.

Low recycling regime with liquid-Li demonstrated the advantage of enhanced confinement, no erosion/blistering of the surface and enhanced plasma purity, which would favour Li liquid walls [Cas 21].

● **Liquid films without pumping effect**

Liquid tin or tin-lithium alloy might be another alternative if massive impurity pumping/gettering could be avoided due to the short pulse length. Such options have demonstrated excellent power handling capabilities in relevant experiments [Dej 20] and present lower vapour pressure than lithium [Cas 20]. Additionally, in the case of the short pulse concept presented herein, the specific concerns related to the high-Z tin impurity ejection under plasma bombardment could be relaxed [Sch 25].

● **Solid coatings for pumping and low recycling**

Solid gettering materials as PFC, like boron or titanium could be conceived. The need of recoating and the flakes are two issues. The potential alternatives are:

**Solid Li coating**. Starting the pulse with a solid lithium coating and allowing it to melt during the pulse would increase the acceptable increment of temperature of lithium.

**Boron coating.** It has low Z and is better for hydrogen pumping than titanium. It would withstand higher temperature than Li without excessive evaporation. Boron is currently being considered as the baseline wall conditioning material to cover the W PFCs in ITER [Wau 25]. The melting point of boron is approximately 2076 °C. The limit of temperature for proper pumping in μStellarators is unclear, but it may be 300–500 ºC depending on the type of boron (amorphous, crystalline) and other conditions. Higher temperature would be favourable.

**Beryllium coating.** It has proved plasma purity in JET, but, low pumping efficiency (it pumps mainly oxygen), toxicity, and the requirement of very low uranium impurities [Kob 16] are concerns.

Other solid gettering materials like **Ti** (higher Z, lower reactivity with H than boron), **Zr-Al** (a getter, Zr has higher Z than Ti), appear to be capable of higher temperatures and could be considered. Nevertheless, they have higher Z and are not well-proven yet for high pumping and low recycling.

Frequent inter-shot recoating (i.e. wall conditioning) is necessary for solid coatings to avoid particle saturation of the coating. In the limit, recoating each pulse could be conceived, but, this is inconvenient for a commercial reactor.

Injection of B, Ti or other material at proper locations for good deposition by the plasma may be implemented, as performed in Ref. [Wau 25].

Elimination of residual flakes, codeposits and debonded coatings may be difficult due to high melting point of such coatings. However, chemical methods have been demonstrated as for example for carbon PFCs [Tab 10], although, they are slow and complex for a commercial reactor.

### 5.B. Materials for the Equi-power heat sink substrate

The heat sink substrate is the layer of material located below the thin film.

A layer of about 5 mm thickness of Equi-power toroidal surface is necessary as heat sink during the pulse. This thickness comes from the expression utilized in [Que 25] for the depth in the layer giving half of the surface temperature. The required thickness slightly depends on the material (Cu, W, Al). The increment of surface temperature ($\Delta T_{surf}$) has been calculated for surface thermal shocks as in [Que 25].

The thin film (liquid or solid coating) is assumed in good thermal contact with the substrate (good wetting).

A thin protective coating between the (likely) lithium and the substrate is necessary for certain substrates, like Cu, Al and alloys, to avoid corrosion of the substrate by lithium while keeping high thermal conductivity. Tungsten, SiC and perhaps alumina are candidates for the protective coating.

**Copper**. Pure copper would be the best conductor among the common materials. However, the thermal conductivity of Cu importantly decreases with neutron fluence. The thermal conductivity of Cu at ~ 0.5 dpa dropped by 10% [Eld 98]. ~ 50% of such loss was recovered by annealing. 0.5 dpa are about 5 operating days at 1% DC. The increase of electrical resistivity (~ thermal conductivity) was considered an exponential function of (-dpa), [Liu 10]. Annealing every few days or continuous self-annealing, if kept at the proper temperature, are potential solutions to radiation damage for a copper substrate.

**CuCrZr alloy** (aged, ~ 0.85 of the thermal conductivity and diffusivity of pure Cu), which is used in ITER divertors, is more radiation resistant than pure Cu. Cu-5Mo alloys showed reduction of less than 20% of electrical conductivity (~ thermal) at ~ 10 dpa of fast neutrons [Fab 98]. However, the effect of thermal neutrons was much higher. 10 dpa are equivalent to ~ 100 operating days at 1% DC. One study reports increase of conductivity of CuCrZr alloy [Eld 98]. Unfortunately, thermal annealing appears difficult or unfeasible due to the compound composition.

**Tungsten** would eliminate the need of the protective coating and could allow the use of the surface of the substrate as capillary support for the lithium, by surface porosity or microchannels on the surface, as i.e. described in Refs. [Evt 99][Ruz 11]. The thermal conductivity of W showed a reduction of 25% with fast fission neutrons at 200 ºC and 0.6 dpa [Roe 04]. The electrical resistivity (~ thermal) increased about 30% for different experiments at 0.9–1.54 dpa of fast neutrons [Fuj 00]. The thermal conductivity halved at about 6 dpa of proton irradiation [Hab 18]. 10 dpa corresponds to ~ 100 days of reactor operation at 1% duty cycle (DC). Annealing seems to recover a relevant fraction of the lost conductivity [Hab 18]. For the Case $A'_{28}$, with ridges and un-irradiated (thermal conductivity at 500 ºC), $\Delta T_{surf}$ ~ 1000 ºC, that would keep the properties of W and might allow the use of Li on W, see Table 2. For the Case $A_{28}$, no ridges and un-irradiated would result in $\Delta T_{surf}$ ~ 2700 ºC ; this temperature would produce recrystallization and W would become fragile at low temperatures –unfeasible W Case $A_{28}$ without ridges. Doped W withstand higher temperatures without recrystallization.

**Aluminium**. Pure aluminium, and alloys with high content of Al, would melt at ~ 660 ºC. Properties of the material will degrade much early. Thus, aluminium hardly could be use as substrate under the studied conditions.

**Static Liquid lithium**. A thick (~ 5 mm) layer of static liquid lithium could be located in an advanced thick CPS. This option appears almost unfeasible even for optim-Cases due to high increase of surface temperature, and thus, excessive evaporation and release of the accumulated hydrogenic isotopes, see Table 1. Thus, moving liquid lithium appears necessary to achieve proper $\Delta T_{surf}$ for some Cases.

Table 2 shows the increment of the surface temperature of the substrate for different Cases and materials (it includes some optim-Cases and one pre-Case). Periodic annealing of the neutron damage is considered, that is, the un-irradiated condition is taken for the calculations. Results with ridges and without ridges are included.

| Material | $\sim \Delta T_{surf}$ (º C). Unirr /Annealed, Ridges. | $\sim \Delta T_{surf}$ (º C). Unirr/ Annealed, No Ridges | Comment |
|---|---|---|---|
| **W** | 1000 ($A'_{28}$)<br>1350 ($A_{28}$) | 2000 ($A'_{28}$)<br>2700 ($A_{28}$) | At ~1200–1400 ºC pure W recrystall. Annealing possible. |
| **Cu** | 450 ($A'_{28}$)<br>700 ($A'_{20}$)<br>420 ($A_{40}$)<br>620 ($A_{28}$)<br>1000 ($A_{20}$)<br>650 ($A''_{40}$) | 900 ($A'_{28}$)<br>1400 ($A'_{20}$)<br>850 ($A_{40}$)<br>1200 ($A_{28}$)<br>2000 ($A_{20}$)<br>1300 ($A''_{40}$) | Annealing at 300–600 ºC, each 10 days (at DC 1%) or constant annealing, is required. |
| **CuCrZr** | ~20% higher than Cu | ~20% higher than Cu | Annealing likely unfeasible |
| **Al** | 850 ($A_{28}$) | 1700 ($A_{28}$) | Melts |
| **Li** | 1600 ($A'_{28}$) | --- | Fast flow needed to decrease $\Delta T_{surf}$ |

**Table 2** Increment of the surface temperature of the substrate for different Cases and materials, with and without ridges. $\sim \Delta T_{surf}$ ≡ approximately, rounded to two significant figures.

As inputs for the calculations in Table 2, the power surface density and exposure time for the Reference Cases is indicated at the beginning of this Section 5. For optim-Cases: $\tau_E = 0.041$ s for $A'_{28}$ ($t_p = 0.20$ s) with pulsed heat load $P_d = 66.5$ MW/m$^2$ for no ridges, and $\tau_E = 0.03$ s for $A'_{20}$ ($t_p = 0.15$ s) with $P_d = 122$ MW/m$^2$ for no ridges. For the pre-Case $A''_{40}$, $P_d = 100$ MW/m$^2$ for no ridges, and $\tau_E = 0.033$ s ($t_p = 0.17$ s). $P_d$ values can be observed/deduced as an approximation from Fig. 7.

### 5.C. Cooling of the Equi-power substrate during off-pulse

One of the main limitations for the reduction of size in toroidal fusion reactors is the extraction of power from ionized particles [Kot 07]. The advantage in μStellarators is the possibility of reduction of the duty cycle, at expense of less fusion power produced.

Let's assume we have a planar copper slab of 5 mm thickness and 24–35 m$^2$ of surface, assimilated to the toroidal Equi-power substrate, corresponding to Case $A_{20}$ and $A_{28}$ respectively. This thickness is able to accumulate the heat from one pulse. The pipes are located at the back of the 5 mm thickness substrate to avoid decreasing the effective surface of the substrate.

The neutronics heat, of less magnitude, is not included next.

Duty cycle 1% and 5% are assessed next.

**Duty cycle 1%.**

For duty cycle 1%, the full internal surface of the μStellarator would have to extract continuously ~ 1 MW/m$^2$ for the Case $A_{28}$, and ~ 2 MW/m$^2$ for the Case $A_{20}$. This comes from the total plasma surface and the full power from alpha particles (beginning of Section 5) and the effect of the duty cycle (1/100). At 1% DC, the electric power produced by the reactor would be ~ 30 MW$_e$ for the Case $A_{28}$ , and ~ 60 MW$_e$ for the Case $A_{20}$ ($\gamma \sim 2$, $P_e = P_{fusion} / 3$ , see Section 2.B, Section 4.B and Fig. 6).

For the Case $A_{28}$ (~ 1 MW/m$^2$), considering pipes of diameter 6 mm at the back of the substrate (embedded in the back), speed of fluid 2.4 m/s, no swirling pipes, 35 m$^2$ of surface (Case $A_{28}$), $\Delta T$ = 25 ºC between the fluid and the internal surface of the pipe, the extracted power would be ~ 35 MW (that is, the total between-pulses-averaged power received by the Equi-power surface for Case $A_{28}$) by using a total length 3000 m of pipes (the number of circuits is not calculated), considering only the convective heat transfer coefficient (in this case conduction through the 5 mm copper slab has low impact). The pipes would occupy about the 50% of the surface in a single layer. This design could be optimised.

**Duty cycle 5%.**

For duty cycle 5%, the substrate of the Equi-power surface would have to extract continuously ~ 5 MW/m$^2$ (Case $A_{28}$) and ~ 10 MW/m$^2$ (Case $A_{20}$). The latter is similar to the ITER divertor targets. For the Case $A_{20}$ , speed of fluid 11 m/s, considering swirling pipes increasing 2-fold the typical heat convection (from ITER calculations [Sal 21]), 24 m$^2$ of surface (Case $A_{20}$), the extracted power would be ~ 240 MW by using 3000 m of pipes. These parameters are similar to the ITER ones. Here, also, the conduction heat transfer coefficient of the Cu slab is not included since copper has very high thermal conductivity. At 5% DC, the electric power produced by the reactor would be ~ 150 MW$_e$ for the Case $A_{28}$ and ~ 300 MW$_e$ for the Case $A_{20}$ (5-fold the 1% DC).

The moving liquid alternatives mentioned in Section 5.D may increase the power extraction, but, at expense of extra complexity related with the free surface flow of fluids.

Higher DC could be envisaged by using microchannels. However, pumping power, the potential clogging of the small microchannels, and extra geometrical complexity under more frequent re-fabrication (higher neutron power) should be studied.

### 5.D. Discussion on power load on divertorial surfaces and materials, difficulties

A thin CPS or a simple coating (re-coated after some pulses) on a ~5 mm Equi-power surface substrate appears the best alternative for μStellarators. The substrate would be made of Cu (or W for certain Cases). The selection may depend on what Cases (A ~ $\beta_{lim}$) can be achieved by plasma physics.

The increase of surface temperature ($\Delta T_{surf}$) of a static lithium layer would be too high (Table 2) and hardly feasible even for optim-Cases.

Tungsten would withstand the pulse (no melting) for the Case $A_{28}$ and some optim-Cases, but recrystallization impose a limit. Nonetheless, W is incompatible with the use of a thin liquid lithium layer on the tungsten for almost all Cases.

Copper would allow the use of liquid lithium for several Cases with ridges (see Table 2), and perhaps for $A_{40}$ and $A'_{28}$ without ridges. Pre-Cases are almost all unfeasible for liquid lithium due to excessive heat load. Optim-Cases, perhaps the Case $A'_{28}$, would relax the conditions and may avoid the need of ridges.

Penetration in the matter of Bremsstrahlung radiation (600 MW for case $A_{28}$) is beneficious, but, not much due to high Z of the substrate.

Lithium is highly corrosive. The surface temperature of the PFC will be continuously (each pulse) oscillating around a relatively high temperature – roughly around 350 ºC. The thin protective coating of tungsten/SiC/alumina on Cu might be corroded too fast. Unfortunately, $\Delta T_{surf}$ is too high for most of the Cases with W substrate, Table 2.

The depth of the porous/micro-grove layer for the thin film cannot be thick to avoid lowering the thermal conductivity of the very first surface of the Equi-power surface. The porous/micro-grove surface should be refurbished periodically due to Li corrosion.

A thin W (~ 0.1 mm) porous/micro-grove layer could act as both protective layer on Cu and CPS, avoiding the need of a protective coating on Cu.

The distribution of ionized particles on the Equi-power surface, as obtained in [Que 25], might still concentrate too much power on individual strike lines (even if there are tens of strike lines), if diffusion of particles is not enough. Diffusive behaviour of particles in non-resonant divertors was studied in Ref. [Bad 17].

Almost the full internal vessel toroid has to be covered by a CPS or coating to achieve $K_d$ = 1 in a Distributed Divertor for a µStellarator design. The positioning accuracy of the full toroidal surface needs to be very high. Indeed, for island divertors (not exactly the Distributed Divertor), the required positioning accuracy is high to avoid hot spots [Fel 23]. The accuracy required for a Distributed Divertor is unknown.

Pumping the helium generated in a µStellarator preferably should be produced by trapping He inside layers of lithium [Sho 23][Mih 24] during the pulse. Alternatively, pumping the helium generated during the pulse could also be (partially) performed during the pulse downtime, if lithium could not trap enough He molecules. The short pulse length implies low burning fraction of the DT (≈ 8% at the end of pulse for Case $A_{28}$) which is favourable for reasonable He ash accumulation.

## 6. Power dissipated in the resistive magnets and temperature increase

Only the power dissipated in the resistive magnets in the flat top of the pulse is studied. The power dissipated during the ramp-up/down, and the power needed to create the magnetic field is out of scope of this work. The later will need recovery and storage in the capacitor banks during the inter-pulse period, in order to achieve net electricity production.

The maximum effective cross section of the coils is pursued by defining thick conducting elements and the full toroidal surface covered by conducting material, in order to minimise the power consumed in the coils. These coils have variable cross section, which is larger at the outboard of the stellarator, as in i-ASTER reactor.

The coils may be located and defined inside the support structure by two **methods/options**:

**1.** As proposed for i-ASTER, that is, without considering much mechanical structure among the coils (only essentially insulation) for maximum effective conductor cross-section, see [Que 18], or

**2.** Coils produced and located by the methods defined and tested in Ref. [Que 22]. Two fabrication methods can be distinguished: **2.A)** Coils of variable cross-section inserted in grooves in contorted-radial plates. The radial plates may be produced by additive manufacturing, as experimentally tested and reported in the work. The conductors may be produced from modified standard filamentary conductors or by 3D-printing, as tested and reported in [Que 22], **2.B)** Turns and layers of conductor with internal insulation, which are wound on a form and subsequent embedded in a low melting point strong alloy, as prototypically produced and reported in [Que 23]. An Inter-cable matrix/structure (Fig. 11) results from both 2.A and 2.B methods.

The space occupied by ports will be negligible due to: **a)** The pulsed regime and the pumping Equi-power surface covering the full in-vessel, **b)** Pulsed regime that allows major pumping between pulses, and, **c)** Focussed ECRH heating for up-to-ignition start-up of the plasma, as already proposed in [Que 10], see Fig. 1. Otherwise, it would be impossible to introduce the heating power in such a tiny device.

### 6.A. Dissipated power in the coils

The method and expressions stablished in [Que 18] are followed (in particular the expression (9)), but, applied to µStellarators.

The pertinent parameters for the µStellarator in this study are $\xi = 1.6$, $\varepsilon = 1$, $f_s = 1.3$, $f_R = 1.2$ and $f_i$, corresponding to µASTER_v1. The coil-shape factor $f_s$ quantifies the increase of length of the coil-turn due to coil twisting. It was obtained in the range of $1.2 < f_s < 1.4$ for QIP3 [Mik 04] and HSR3 magnetic configurations. We assume that the large aspect ratio magnetic configuration for µStellarators has $f_s$ at the middle of the observed range. $f_R$ quantifies the increase of the length of the magnetic axis in the stellarator in relation to that of a (circular) tokamak magnetic axis. $f_i$ is the ratio of conductor cross-section $S_{Conductor}$ to the total section available.

The electrical resistivity of aluminium is taken $\rho = 4.1 \times 10^{-8}\ \Omega$ m at 150 ºC and the volume-specific heat of aluminium $C_p = 2.45 \times 10^6$ J/(m$^3$ K). 150 ºC is taken as an intermediate temperature of the conducting material from ~ 50ºC initial temperature to the final temperature, which depends on each Case. Cases $A_{40}$ and $A_{28}$ would be approximately inside this range. Case $A_{20}$ would give higher temperature rise and the estimation should be adjusted.

For copper $\rho$ = 2.55 ×10$^{-8}$ Ω m at 150 ºC. For copper $f_i$ = 0.6 (= 0.95 * 2.55/4.1) to obtain the same resistance of the coils as for aluminium. This would give superior structural support of the cables in the coils. Copper would allow the constructive method '2'.

Variable resistivity depending on the temperature during the pulse is not taken into account.

Fig. 8 represents the approximate electric power consumed in the resistive coils for the considered parameters and the minimum field for ignition presented in Fig. 5.

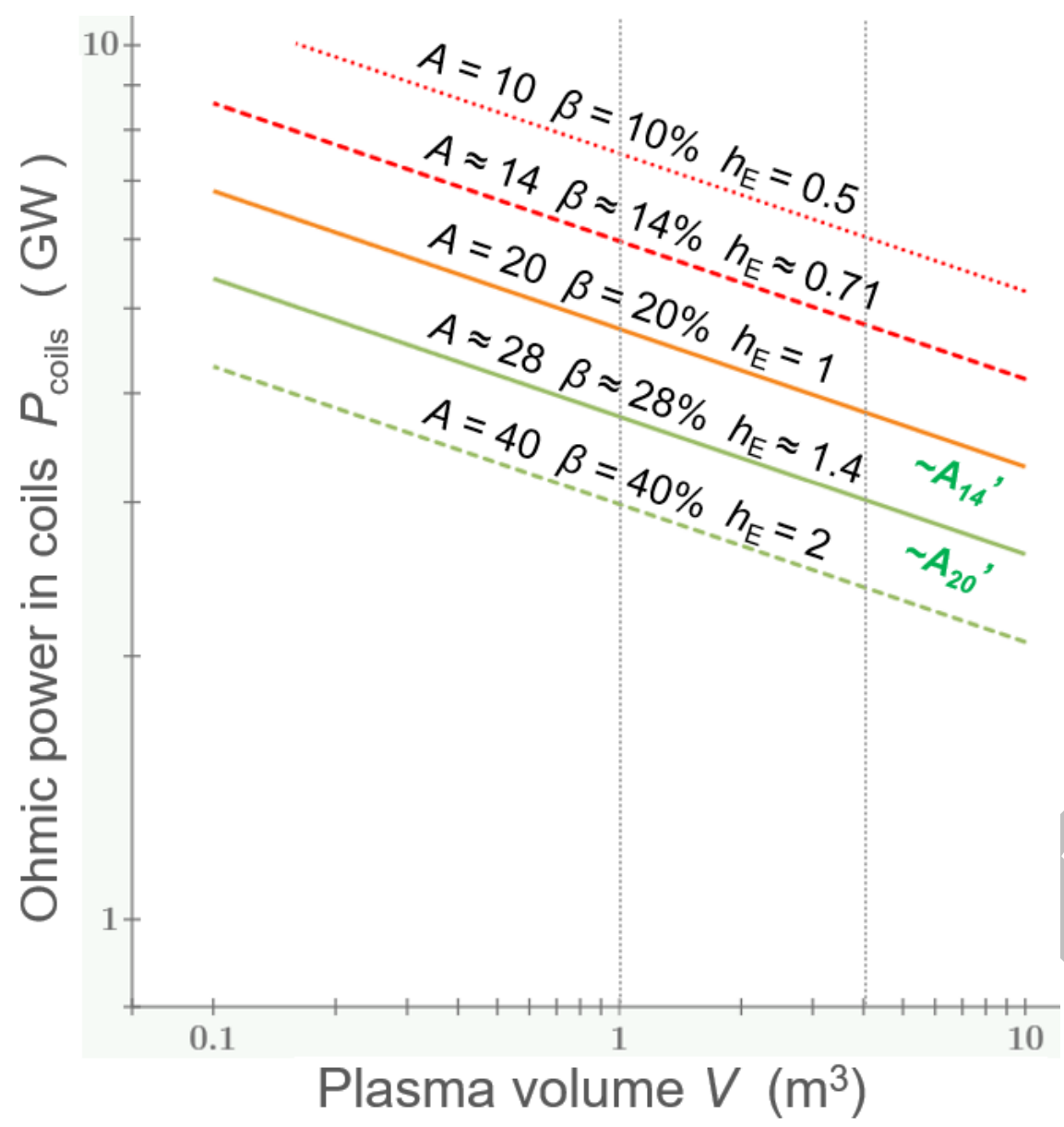

**Fig. 8** Approximate electric power consumed in the resistive coils for the coil parameters in Section 6.A, and $f_i$ = 0.95 for aluminium and $f_i$ = 0.6 for copper, and the magnetic field for ignition presented in Fig. 5. The optim/pre-Cases does not coincide with the lines for the Reference Cases. Two optim-Cases are represented where approximately they would have the plot-line.

***Discussion on dissipated power in coils, difficulties***:

The power consumed by the coils is 2.4, 3.2 and 4.2 GW$_e$ respectively for the Cases A$_{40}$, A$_{28}$ and A$_{20}$ . The other Reference Cases are likely unfeasible.

For aluminium, $f_i$ = 0.95, corresponding to the alternative '1', has little inter-cable matrix/structure and the mechanical strength is compromised. $f_i$ = 0.6 for copper would allow considerable inter-cable matrix, Fig. 11.

### 6.B. Coil temperature increase

The method and expressions stablished in [Que 18], in particular the expression (10), is applied to the µASTER_v1 parameters.

The maximum increase of temperature at certain points at the coils resulted

$$\Delta T_{max} = \Delta T_{ave}\ \ f_c{}^2$$

being $f_c$ the concentration factor for the maximum current density relative to the average current density, [Que 18]. $f_c$ = 5 for QIP3 and $f_c$ = 6 for HSR3 was calculated [Que 18].

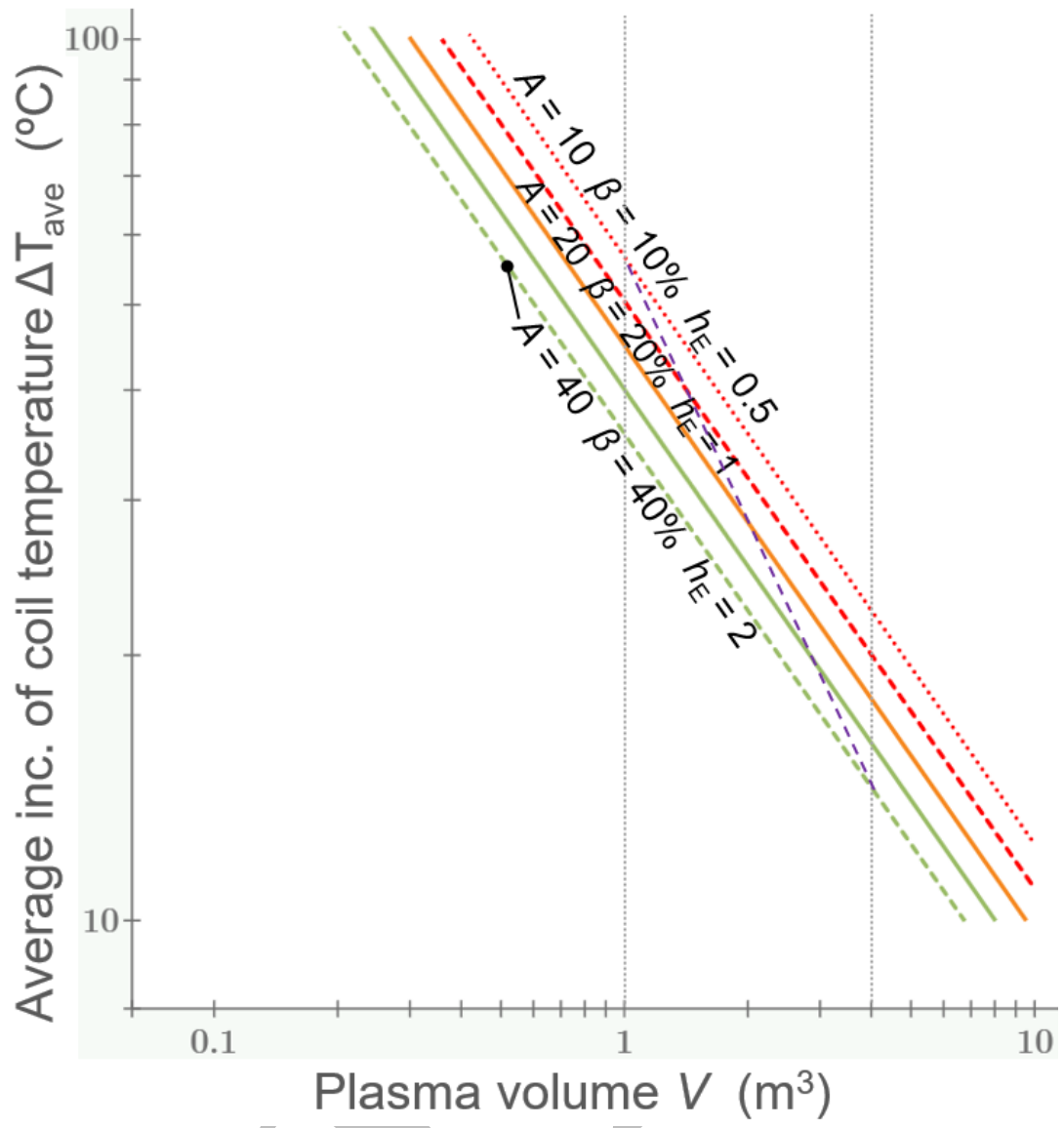

**Fig. 9** Average increment of temperature $\Delta T_{ave}$ of resistive aluminium coils for the parameters in Section 6.A and the magnetic field for ignition presented Fig. 5. The violet line crossing the different cases indicates devices with $a$ = 0.17 m. The parameters for the unlabelled lines correspond to the respective ones in the other plots.

***Discussion on coil temperature increase, difficulties:***

Most of the segments of aluminium or copper conductors would increase their temperature around the average increase $\Delta T_{ave}$. However, $\Delta T$ at the coil segments located at the inboard curved sectors of the stellarator would be $\Delta T_{ave}\ f_c{}^2$. Thus, reducing $f_c$ as much as possible is paramount. As cited, $f_c{}^2$ was calculated in the range 25–36 for two magnetic configurations. $f_c$ = 4 or lower would be convenient.

The variation of $\Delta T_{ave}$ in the interval 1–4 m$^3$ is the largest among all the other parameters studied. Thus, increasing plasma volume is very relevant to have lower $\Delta T_{ave}$ .

For example, if we would consider $f_c{}^2$ = 16 and $\beta_{\text{lim}}$ = 28% (Case A$_{28}$) it results in $\Delta T_{ave} \approx$ 19 ºC and $\Delta T_{max} \approx$ 304 ºC. This increment of temperature might be possible for aluminium, though, it is notably high. The mineral insulation appears capable to withstand such increment of temperature since some Mineral Insulated Cables (MIC) are used as heaters under radiation, reaching up to ~1000 ºC. For Case A$_{20}$ ($\beta_{\text{lim}}$= 20%) the increase of temperature appears hardly achievable for aluminium, but, appears feasible for copper.

### 6.C. General discussion on the resistive magnets

The design and construction of magnets of such thickness ($\varepsilon$ = 1) appears difficult, though additive manufacturing and other advanced methods could help the fabrication of thick layer(s) of conductor and insulation, see i.e. [Que 22].

The increase of electric resistance due to radiation seems less problematic than the neutron degradation of the coil support, since the conductor could be annealed/re-melted/purified in-site by using the inter-cable matrix as a mould. Electrical conductivity of copper seems less affected by neutron radiation while aluminium appears more sensitive, see Section 8.B. The reuse of the insulation after neutron damage appears more difficult, due to the likely ceramic/mineral composition.

The possible non-uniform increase of the conductor temperature (~resistivity) if the coil is thick (one or few layers of turns) needs further study.

## 7. Estimation of stresses in the coil support

An estimation of the stress in the coil support, the study of the properties of potential materials for the coil support and conductors, mechanical fatigue, the selection of possible materials for the coil support and a final discussion is included in this section.

### 7.A. Estimation of the stresses in the coil support

An estimation of the stress in the monolithic coil support is performed in this section. The maximum achievable magnetic field $B$ is limited by the yield strength of the material of the monolithic coil support, of the inter-cable matrix/structure and the insulation, under the working conditions.

The method and expressions stablished in [Que 18], in particular the equation (12), is applied to the µStellarator, Fig. 10. FEA calculation of stresses at small locations (i.e. due to the effect of each radial-plate-like coil at the high field inboard areas of the monolithic coil support) was produced for i-ASTER [Que 20], but, not for the µStellarator.

Fig. 10 shows the approximate average stress $\sigma_s$, as defined in [Que 18], in the monolithic coil support for $\psi = 0.75$, $\xi = 1.6$ and the other parameters in Section 3 and 4.

The maximum stress in the structure is $\sigma_{max} = f_\sigma \, \sigma_s$, where $f_\sigma$ is a stress concentration factor. Calculations performed for QIP3 resulted in $f_\sigma \sim 2.5$ [Que 18]. $f_\sigma$ varies with the aspect ratio and the level of convolution of the stellarator. A high aspect ratio tokamak would have $f_\sigma \rightarrow 1$, since the maximum and minimum magnetic field tend to the average value and the current density on the winding surface (circularly toroidal in a tokamak) is almost constant on the surface. The aspect ratio in i-ASTER is $A$=6 while the aspect ratio of the Cases $A_{20\text{-}40}$ in the present study is much larger than that. Additionally, the mirror ratio $B_{max}/B_{min}$ on the LCFS in low number of periods common QI magnetic configurations is much higher than for high aspect ratio [Mik 04]. For example, the QIP3 configuration, with $A \approx 7$, has mirror ratio of 2.9 [Mik 04]. QIP9, with $A$ = 24, has a mirror ratio of 1.5 [Mik 04]. Defining $\delta B_{axis} = B_{max,axis}/B_{average,axis}$, $\delta B_{axis\text{-}QIP3} \approx 1.73$ and $\delta B_{axis_\mu Stell\text{-}A20} = 1.25$ (that is, for the $A$ = 20 configuration in Section 3.B).

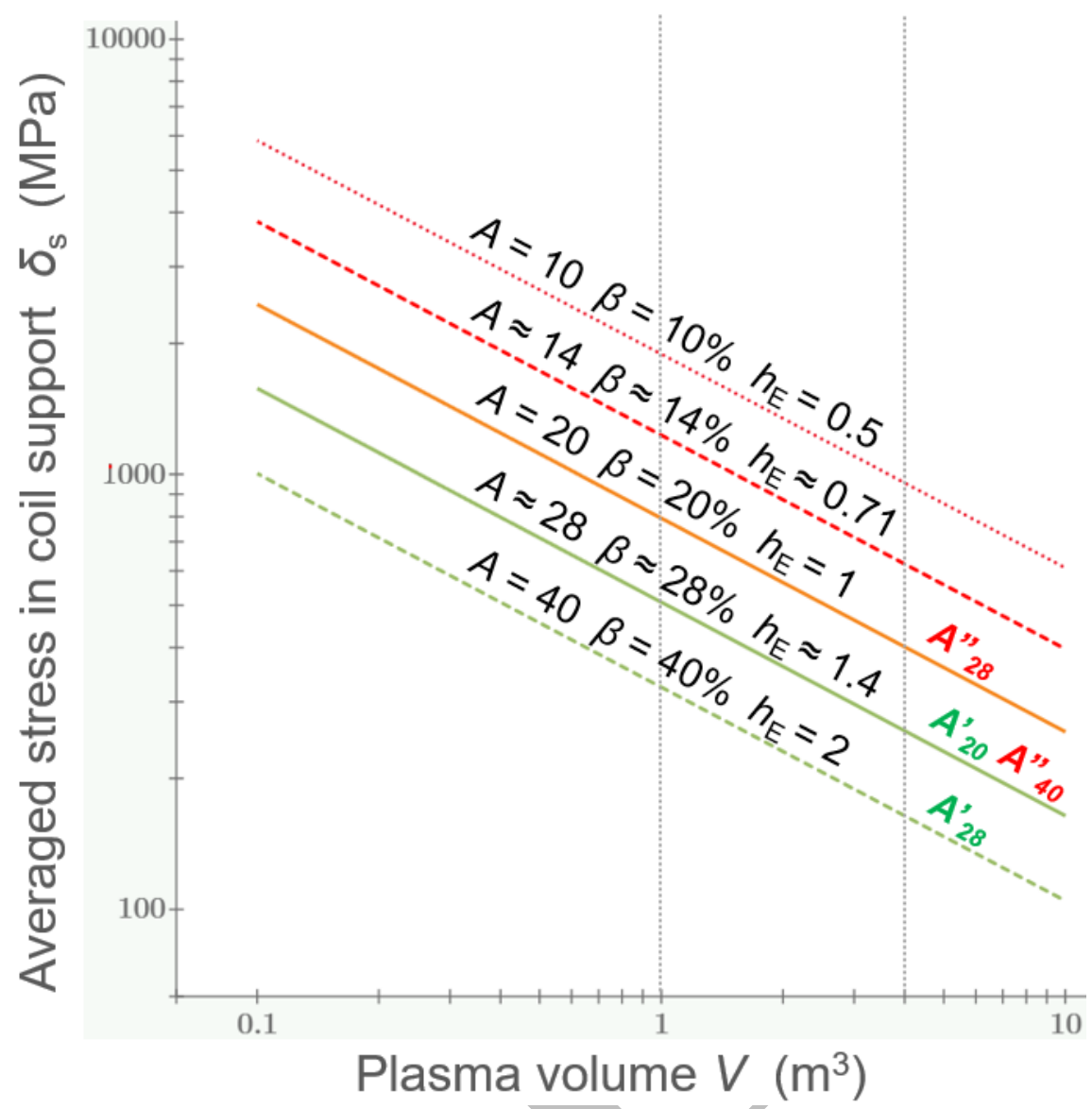

**Fig. 10** Approximate average stress $\sigma_s$ in the monolithic coil support for $\psi$ = 0.75 (thickness = 0.75 $a$) for the magnetic field for ignition presented in Fig. 5. The optim/pre-Cases agree with the proper line of a Reference Case.

Defining $\delta B_{out\text{-}in} = B_{outboard}/B_{inboard}$, $\delta B_{out\text{-}in} \rightarrow 1$ for large aspect ratios and common magnetic configurations, since the curvature radius tends to $\infty$ ($\nabla B \propto 1/R_C$, being $R_C$ the radius of curvature of the magnetic field lines). For the sectors of maximum curvature, $\delta B_{out\text{-}in_QIP3} \approx 1.23$ and $\delta B_{out\text{-}in_\mu Stell\text{-}A20} \approx 1.2$, obtained from the magnetic fields. For QIP3 the product of the two factors $\delta B_{axis} \times \delta B_{out\text{-}in} = 2.1$, and for µStellarator-$A_{20}$ = 1.5. Thus, as a first approximation $f_\sigma$ might be $f_\sigma \sim 1.8$ for the µStellarator-$A_{20}$ (1.5 × 2.5/2.1). For Case $A_{28}$ it should be slightly lower, and we assume $f_\sigma = 1.6$.

Large aspect ratio stellarators seem to moderate the maximum stresses in the coil support, which is important.

For $\psi$ = 0.75, the average stress in the coil support is $\sigma_s \approx 160$ MPa for the Case $\mathbf{A_{40}}$ (and ~$A'_{28}$ due to small change in plasma volume) and $\sigma_s \approx 300$ MPa for $\mathbf{A_{28}}$ (and ~$A'_{20}$). And the maximum stresses are expected **$\sigma_{max}$ ~ 250 MPa and 480 MPa** respectively, considering $f_\sigma = 1.6$ for both. Case $\mathbf{A_{20}}$ gives **$\sigma_{max} \approx$ 900 MPa**, which is rather high, even for $f_\sigma = 1.6$.

Low activation materials should be preferably considered, which reduces the choices and the maximum strength of the materials.

Also, mechanical fatigue is an important factor of reduction of strength of the materials in the present pulsed reactor. It is assessed in Section 7.E.

### 7.B. Properties of potential materials for the coil support

Table 3 shows some steels and a Zr alloy, for the selection of a potential candidate for the monolithic coil support. Strength at Room Temperature (RT) and 200ºC, level of low activation, neutron transparency, resistance to neutron swelling and general neutron resistance are assessed. The intention is to compare the different materials for an initial balancing of the properties, not to have accurate values for very specific materials.

Some elements may be favourable with isotopic purification, to achieve enough TBR without reducing the plasma size and/or $\psi$.

The increment of temperature at the internal surface of the coil support at peak neutron load regions will be roughly 120 ºC during the pulse (Section 8.E), though it depends on the material. Considering initial temperature 50 ºC and certain margin for the spatial variations, 200 ºC is considered an initial reasonable assumption for the end-of-pulse coil support temperature.

The coil support materials would be reused/re-melted, which reduce the mass/volume of wastes. Still, the activation (in Bq) for most of the long-lived isotopes increases with time (many remeltings ~ operation time). Some data for Mo, Co, Ni and Zr is summarised next and comes from Ref. [Gil 15].

Mo is totally unacceptable as an alloying element due high long term activation ; $\sim 10^4$x LT Bq/γ/decay heat.

Nb is also not acceptable as an alloy component since the activation values are ~100x LT Bq ; $10^6$x LT γ [Gil 15]. Thus, Zr alloys like Zr-2.5Nb (used in CANDU fission reactors) and Zr-Nb-Ti-with-nano-precipitates are not included in the table due to the high activation.

Co and Ni suffer also from high activation, but lower than Mo. The specific activity (Bq) of Co is two orders of magnitude lower than Ni in the long term >1000 years. However, the Bq/γ/decay heat during the first year is 1-2 orders of magnitude higher in Co than in Mo and Ni, which is more troublesome for remote maintenance and decay heat removal. Co has ~ 1x Bq , 100x γ/decay ~300 years after shutdown. Thus, concerning rad-wastes a small alloying proportion of Co may be utilized.

Zr-nat has similar activation (Bq/γ/decay) to Ni in the long term (1000 years). Advantageously, from 1-100 years the activation is low, lower than Fe. Isotopic purification of Zr to use only $^{90}$Zr will much reduce activation of Zr, see Section 8.F.

ODS V-4Cr-4Ti alloy is not included due to poor neutron transparency of vanadium.

Ni/Co/Mo maraging steel is included for comparison, due to its very high tensile strength, though it is not appropriate for a commercial reactor due to high activation and poor performance under neutron radiation (rapid embrittlement, dissolution of precipitates, high swelling).

| Alloy | ~$S_y$ RT (MPa) | ~ $S_y$ 200ºC (MPa) | Non-Activation | n. trans. | Resist. to n. swell | n. resist | Ref. / Comment |
|---|---|---|---|---|---|---|---|
| Fe-9Cr-W-V-Ta precip. hardening (alike EUROFER 97, F82H, CLAM) | 530 [Fer 01] | 490 [Fer 01] | H | H: $^{54}$Fe removed M: Fe-nat | H [Che 13] [Tod 18] | H* [Pet 06] [Jit 09] | Nuclear qualified. * Suitable neutron resistance depends on irradiation temperature (DBTT). |
| **ODS Fe-13-14Cr-2W-0.3-0.4Ti-0.3x $Y_2O_3$** **ODS Fe-14Cr-0.3$Y_2O_3$** **ODS Fe-9Cr-0.35$Y_2O_3$** | 1430 [Yan 13] 1000 tempe-red [Yan 13], 1070 [Zha 23] 1100 [Aug 13] (HIP) | 950 [Aug 13] | H | H: $^{54}$Fe removed. M: Fe-nat | H* | H* | * Assumed similar to EUROFER. |
| Fe-18Ni-9Co-5Mo maraging steel | 1900, DMLS [Pro 26] 1870 [Yir 23] | | L | --- | L | L | Not suitable for a commercial reactor due to neutron damage and high activation. Relatively simple fabrication. Ductile tearing. Age hardened. |
| **ODS 49Zr-30Fe-10Cr-5Cu-5Ti-1$Y_2O_3$** | 1730–2550 * | Likely able for high T | H, if $^{65}$Cu and $^{90}$Zr | H | --- | --- | Pulse plasma sintering, alloy B, [Kar 18]. Lack of data on neutronics effects. * Compressive strength; value depending on sintering T [Kar 18] |

**Table 3** Some steel alloys and a Zr alloy for the selection of potential candidates for the monolithic coil support and inter-cable matrix. In bold letter the best candidates. Strength at RT and 200ºC, level of low activation, neutron transparency for tritium production, resistance to neutron swelling and general neutron resistance are included.
L ≡ Low ; M ≡ Middle ; H ≡ High.
Non-activation: Low activation materials have "H" high performance. Data comes from the First Wall values in [Gil 15].
n. trans = Neutron transparency for tritium production.
n. swell = Resistance to neutron swelling.
n. resist = Resistance to neutron damage on the strength (strength decrease and ductile-to-brittle transition temperature (DBTT)).

## 7.C. Properties of possible materials for the conductors

The mechanical strength of the conductor is also important, though generally it will not have key structural function.

The yield strength $S_y$ of the aluminium alloy 6061-T6 (Al-Mg-Si-Cu) is ~ 310 MPa at RT and ~ 120–150 MPa at 200 ºC. Precipitate coarsening (β"-$Mg_2Si$) begins at 200 ºC. Electrical conductivity is ~40–43% IACS at RT, and ~30–35% IACS at 300 ºC, un-irradiated (pure Al has 62% IACS at RT). Other stronger Al alloys like the 2618A (Al-Cu-Mg-Fe-Ni) have too low electrical conductivity for a reactor with net-electricity production, ~30–35% IACS at RT and un-irradiated.

$S_y$ of pure aluminium is ~20–30 MPa at 200 ºC which is very low. It might be used only for an innovative design alternative.

$S_y$ of the copper alloy Cu-Cr-Zr (C18150, cold-worked, aged) is ~400–500 MPa at RT and ~250–300 MPa at 300°C. Electrical conductivity is ~80–85% IACS at RT and ~70–75% IACS at 300°C (un-irradiated; compared to 100% IACS for pure Cu at RT). Lower alloyed C15000 (Cu-0.1Cr-0.06Zr) has $\sigma_y$ 200–250 MPa at RT and ~90–95% IACS.

Other stronger Cu alloys like beryllium copper (C17200, age-hardened, $S_y$ ~400–500 MPa at 300 ºC) have too low electrical conductivity for net-electricity production, ~22–28% IACS at RT and un-irradiated.

### 7.D. Properties of the mineral insulation

MIC cables are the only possibility for the μStellarator coils due to the high radiation (organic insulation is unfeasible) [Kal 11]. The ITER In-Vessel coils [Kal 11] are the reference for the mineral insulation material and geometry in μASTER_v1. Indeed, this technology was already used in the 70's [Har 70].

The selected oxide for the mineral insulation is MgO since it showed good performance under different types of loadings [Kal 11], and it was possible to produce joints and to compact the MgO at the joints [Yin 23]. The low shear strength of organic insulators seems less problematic for MICs. Indeed, the transmission of shear loads from the external metallic sheath/sleeve to the internal conductor Cu/Al seem not required in MICs [Kal 11]. Still, compression loads have to be withstood, though, the option of numerous turns (Fig. 11) would reduce these normal loads in μStellarators.

These in-vessel ITER coils will have to support high electromagnetic and thermal loads [Vos 19], and the ELM coils will suffer from important mechanical fatigue.

The thickness of the MIC insulation for ITER IVCs is 2.5 and 5 mm [Kal 11], about 10 times lower than the diameter of the conductors. Thus, high effective Cu/Al cross section is foreseen for μStellarators. Differently, the effective cross section of Cu was modest in the cables in Ref. [Har 70], only ~ 50%.

Undoubtedly, further studies and tests for μStellarator conditions are required.

### 7.E. Mechanical fatigue

Pulsed devices like the μStellarator have to cope with mechanical fatigue.

Mechanical fatigue is a complex mechanical matter that may require testing even after performing many analytical and FEA calculations, as happens i.e. in aerospace technology.

High quality design, proper surface treatment, appropriate manufacturing, tests, damage-tolerant design and periodical inspections will be necessary for μStellarators.

An advantage in μStellarators, as happens in i.e. i-ASTER, is that the highest stresses are compressive [Que 21] and stresses are non-reversal (does not change from tension to compression). This will improve the Fatigue Ratio (FR = $S_\infty/S_u$). It is expected that the general behaviour of the stresses (i.e. type of principal stress in each area) in μStellarators will be similar to i-ASTER in Ref. [Que 21], since the mechanical and magnetic configuration is similar. The Fatigue Ratio for full reversal is in the interval 0.35–0.45 for most of the materials in Table 3. A Fatigue Ratio of 0.7 for non-reversal stresses is assumed here to start to assess the matter.

### 7.F. Selection of possible materials for the coil support

**Case $A_{28}$.** From Section 7.A, for the Case $A_{28}$ (and approx. for **Case $A'_{20}$**) a material having $S_y \approx 700$ MPa ($\sigma_{max} \approx 480$ MPa) would be required when considering non-reversal fatigue (fatigue ratio 0.7), for $\psi = 0.75$.

Two of the compiled materials may reach such strength at 200 ºC, are neutron-transparent enough, and may be appropriate for the neutron environment: ODS Fe-14Cr-3W-0.3Ti 0.3$Y_2O_3$ and ODS 49Zr-30Fe-10Cr-5Cu-5Ti-1$Y_2O_3$. Depending on the heat treatment and particularities, some of them could be appropriate for $\psi = 0.5$. Indeed, reduction of the thickness of the coil support would increase neutron transparency.

Thus, at least some materials exist for the physics-reasonable Cases $A_{28}$ and $A'_{20}$. Nevertheless, only the $\beta_{lim}$ of Case $A''_{20}$ has been currently achieved. Thus, further improvement of physics for $A = 20$ and obtaining one configuration for the Case $A_{28}$ will be important.

**Case $A_{40}$.** For the Case $A_{40}$ a material with $S_y \approx 350$ MPa ($\sigma_{max} \approx 250$ MPa ; FR=0.7) would suffice. There are many low activation materials that would withstand such stresses under fatigue, included EUROFER97, F82H or equivalent RAFM steels, Table 3. Also, common (non-ODS) Zr Alloy 1wt% $Y_2O_3$ may suffice. However, $A_{40}$ is hardly physics-feasible due to the high beta limit involved.

**Case $A_{20}$** (and ~**$A'_{14}$**). $S_y \approx 1300$ MPa is required to withstand fatigue. Thus, only maraging steels (and perhaps ODS 49Zr-30Fe-10Cr-5Cu-5Ti-1$Y_2O_3$) would cope with the high stresses and fatigue, see Table 3. Maraging steels may only be used for an experimental reactor, that is, having a modest number of pulses, to limit the neutron damage and swelling. IFMIF-DONES facility [Kan 15][Ber 22] should improve the radiation resistance of high-strength steels and Zr alloys, and thus, might find other material alternatives for Case $A_{20}$ and $A'_{14}$ μStellarators.

### 7.G. General discussion on mechanical stress and materials, difficulties

The stresses are huge for Cases of lower aspect ratio and smallest plasma volume. These Cases ($A_{10}$, $A_{14}$) are unfeasible.

Case $A_{28}$ and $A_{40}$ appears potentially feasible concerning stresses, and using low activation and low swelling materials. Case $A_{20}$ has more limited material alternatives.

Swelling is an important matter in this application since high accuracy is necessary to keep the precision of the magnetic configuration.

FEA calculations for at least one Case would substantiate the concept. General expressions, adjusted by factors, are appropriate for the initial development of a concept, but, particularisation would be necessary in the future.

FEA studies on the effect of the individual cable and conductor in the inter-cable matrix are not produced. Also, the particular design for piling or embedding the cables in each coil will have impact on the stresses.

Fabrication methods for the coils and supports were studied and tested in Refs. [Que 22][Que 23], but, the tests are still preliminary.

The relatively low strength of suitable Cu alloys for the conductor, and rather low Al alloys (Section 7.C), require moderate stresses in the conductor from electromagnetic forces. One potential solution is to use a considerable number of turns per coil. This divides the magnetic forces while keeping proportionally higher surface of conductor. One alternative for implementation of the concept implies embedding the sheathed cables in a strong inter-cable matrix, as proposed and tested in [Que 23], Fig. 11. An alternative could be the use of contorted radial plates, somewhat as used in ITER (planar plates in ITER case) [Que 22]. Another potential option could be the use of radial plates working by lateral compression due to the centring forces, somewhat as performed in Alcator C-Mod.

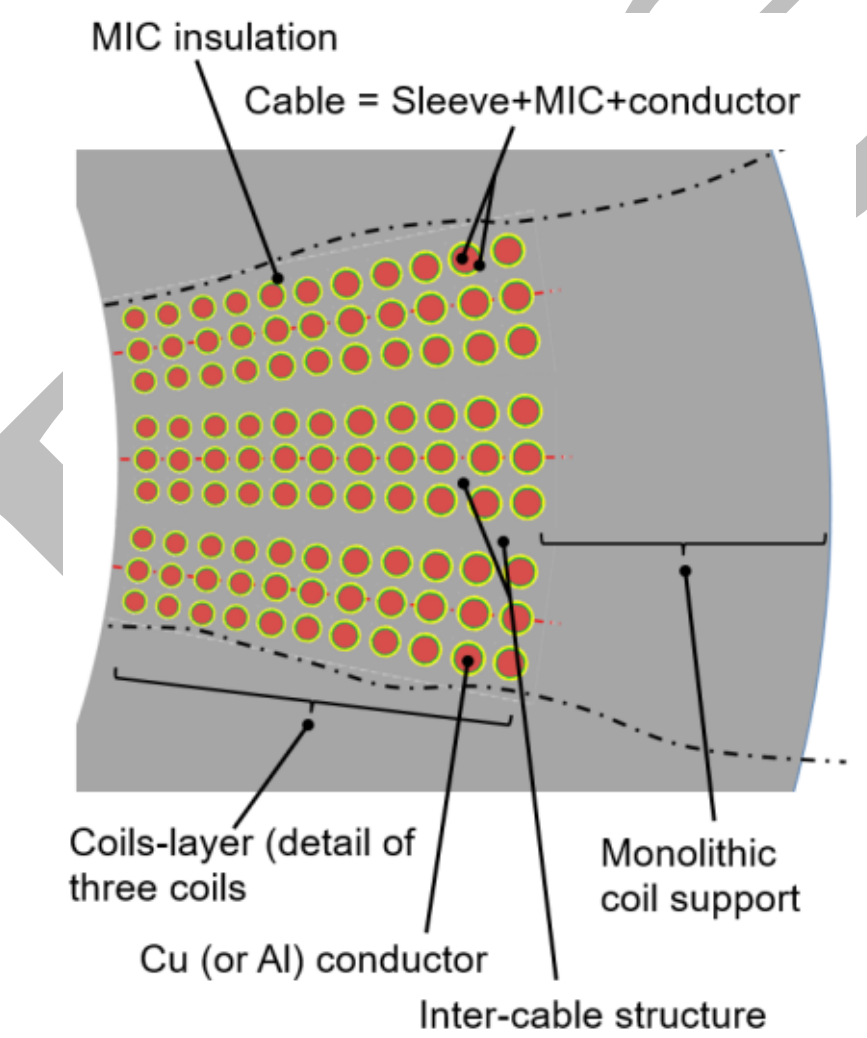

**Fig. 11** Concept of numerous conductors in an Inter-cable structure/matrix or contorted stellarator radial plates. See further information in [Que 22][Que 23].

Mechanical fatigue increases the difficulty of the concept. Also, stresses due to thermal cycling are important. Insulation failure occurred in NSTX and JT-60SA tokamaks. Indeed, the failure rate of coils should be very low in any commercial reactor. Favourably, the half-periods of the μStellarator would be splittable, similarly to defined in [Que 18]. Detachable half-periods are relatively simple for resistive coil reactors and more difficult for superconducting coil reactors [Pal 26]. Thus, in μStellarators the repair would (only) imply disassembling a half-period, replace it, and off-line repair/re-fabrication of the failed one. Certainly, in any case, the half-period has to be refurbished or re-fabricated periodically due to neutron damage.

Deformation of the structure, due to the Lorentz forces might be acceptable, as deduced from [Que 20] for magnetic fields somewhat higher than $B_0$ = 10 T and for steel. The deduction is imprecise since the aspect ratios (i-ASTER and μStellarators) are rather different.

Methods for the production of the monolithic coil support were studied, developed and assayed by the authors in the past [Que 15,16,21,22,23]. Either contorted radial plates or embedding the conductors in a matrix are the preferred options.

## 8. Neutronics: TBR. Lifetime. Decay heat. Activation.

The Case $A_{28}$ can be considered as a middle point of reference for the neutron effects. The Case $A_{28}$ produces pulsed fusion power $P_f$ = 18 $GW_{th}$, neutron power 14.4 GW during $t_p$ = 0.16 s, and has ≈35 $m^2$ of plasma surface (considered equal to the vacuum vessel and Equi-power surface), resulting in ~400 $MW/m^2$ of pulsed neutron power density.

The materials are considered pure in the sense of lacking of long-term high-activation impurities, i.e. Nb, Ag. This is different to having a moderate activation element as an alloying compound.

TBR, radiation effects, decay heat, neutron heating, lifetime and activation are estimated for the main elements of the reactor core.

### 8.A. Neutron transparency of the shell for TBR

The next calculations are performed with MCNP5 v1.6.

A simpler spherical model is considered instead of the more realistic toroidal model for this initial evaluation. Next, all the calculations are performed for a spherical model.

Isotopic purified elements are utilized in some of the next shells in order to increase the TBR (and avoid the use of Be multiplier) and much decrease the activation of certain elements (i.e. Cu, Zr).

Concerning steel alloys, the neutron transparency for T production of $^{52}Cr$ is similar to Fe-nat, thus, pure Fe-nat or proper Fe isotopes can be utilized in the TBR calculations instead of the Fe-Cr alloys. It is conservative and more general. Ti could be used in lower alloying proportion than Cr, and ~1%wt of W seems not significant for TBR. Thus, only pure Fe is used in the first of the next three calculations of shells.

Concerning zirconium material, the neutron transparency of $^{90}Zr$ is similar to Zr-nat. Thus, the option of using $^{90}Zr$ to avoid most of the Zr activation does not hinder the neutron transparency of Zr (is conservative).

**$^{65}$Cu conductor and $^{57}$Fe coil support and matrix.** $^{65}$Cu is utilized to much reduce activation and increase neutron transparency. $^{57}$Fe is used for its higher n-transparency. For copper conductor ~40% of the Coils-layer is made of matrix and 60% of copper, Fig. 8, Section 6.A. This corresponds to 0.102 m (0.17*0.6) thickness of Cu, and 0.153 m of Fe (0.068m inter-cable matrix + 0.085 coil support). Note that here $\delta \approx 0.255$ m, $\psi = 0.5$ , a = 0.17 m. A sphere of internal diameter ~10 m with the proper thickness of the two layers and materials is modelled in MCNP5. Also, there are 2 m thickness of natural lithium externally. It results **TBR = 1.0**, which is considered enough for this initial evaluation.

**$^{65}$Cu conductor and $^{90}$Zr coil support and matrix.** $^{90}$Zr is utilized to much reduce activation. The copper layer is the same as in the previous case (0.102 m). The $^{90}$Zr layer is 0.196 m thick (0.068 m inter-cable matrix + 0.128 m coil support). Note that here $\delta \approx 0.3$ m, $\psi = 0.75$ , a = 0.17 m. The same sphere of internal diameter ~10 m with the proper thickness and materials, and the same Li-nat is modelled. It results **TBR ≈ 1.1**, which is considered enough.

**Al conductor and Zr-nat coil support and matrix.** For these materials and $\delta \approx 0.3$ m, 0.05 m of beryllium multiplier (facing the plasma) is required to achieve TBR near 1 [Que 25b]. This option is not further studied, since, additionally, the inter-cable matrix would be rather weak.

### 8.B. Radiation lifetime of metals of the reactor core

The monolithic coil support and the conductor material of the coils are included in this section. Materials for electrical insulation, which are also located inside the reactor core, are studied in Section 8.C.

**Monolithic coil support and inter-cable matrix.**

For duty cycle of 1 % it results in ~ 40 dpa/year (400 MW/m$^2$ × 0.01 DC × 10 dpa/(MW-y/m$^2$), assuming a full power year of 1 MW-y/m$^2$ ≡ 10 dpa). The potential extra effects of being high power pulses are considered negligible. Considering 50 dpa the limit of radiation damage for a steel (Table 3) monolithic coil support and inter-cable matrix, the reactor core will require replacement about each year (that is, lifetime about one year). This result for the µStellarator is coherent with the replacement frequency of blankets in typical stellarator reactors. Indeed, at ~1 MW/m$^2$ in almost continuous work, blanket replacement is planned each 8 yr in HSR22 and 10 yr in FFHR [Ama 99][Sag 05]. Fluence limit ~ 100 dpa was considered for steel in such model of HSR reactor [Ama 99] and FFHR [Sag 05].

Zirconium alloys used in most fission reactors showed important hardening at high neutron doses [Oni 20]. However, ODS Zr alloys with nano-precipitates (Table 3) may be more neutron resistant.

The simple top part of the support legs will also require replacement due to radiation damage.

**Aluminium conductor.**

The electrical resistivity of the aluminium conductor increases with radiation damage due to the accumulation of radiation defects and transmutation. For the aluminium alloy SAV-1 (Al 97.6wt%, Si 2.1wt and others), 0.1 dpa increased the resistivity 10% [San 11]. The possible saturation of the increase of resistivity of aluminium for higher radiation doses (i.e. > 1 dpa) need to be further studied. It appears to occur with copper [Liu 10].

Volumetric (density based) swelling of hardened 6061 aluminium alloy was 0.1% for neutron fluence of 3 x 10$^{22}$ neutrons/m$^2$, up to 3% for other Al alloys, and higher for high purity Al [Far 73].

For neutron fluence of ~10$^{23}$ neutrons/m$^2$, the yield and ultimate strength of aluminium increased with radiation about 50% at 50 ºC and does not changed significantly at about 200 ºC [Far 73].

**Copper conductor.**

The increase of electrical resistivity is considered an exponential function of (-dpa) in Ref. [Liu 10]. This indicates a positive behaviour of the conductivity of copper with increasing radiation. Copper has higher conductivity than aluminium and allows much more space for the inter-cable matrix.

### 8.C. Radiation effects on the electrical insulation

Potential issues due to neutron/gamma radiation on electrical insulation of conductors are assessed next. Here, an insulation for power coils is considered (not for instrumentation insulation).

Only fully mineral insulation is considered, due to the unfeasible organic insulation at such high radiation dose, Section 7.D. The insulation has to be in the form of powder or fibres, since the mechanical strength and toughness of bulk ceramics is too low for this application, even without irradiation. Thus, the reference insulation for µStellarators is a Mineral Insulated Cable, Section 7.D.

The most relevant phenomena generated by radiation for this application are evaluated next.

*Radiation induced conductivity* (RIC) appears negligible for this application. For example, electrical conductivity of alumina increased to ~10$^{-6}$ S/m for ionizing dose rate two orders of magnitude higher (~ 10$^6$ Gy/s) than in the first wall of typical fusion reactors (~ 10$^4$ Gy/s) [Neu 10]. Also, this data suggests no RIC issue for MgO and $MgAl_2O_4$ (spinel) insulator.

*Electrical degradation*:

MgO decreased the **resistivity** a factor 3–4 at 10 dpa of neutron irradiation [ITE 97]. Other neutron irradiations are currently being produced [Sha 24]. The value suggests that electrical degradation of the insulation, due to the cumulative neutron damage, will not be critical for this application (other factors will prevail for the replacement of the reactor core).

Measurements indicate that the **dielectric strength** of certain ceramics decreased by a factor of 2 to 4 under high dose neutron irradiation [ITE 97], which has to be taken into account.

*Mechanical degradation due to radiation:*

Toughness of spinel ($MgAl_2O_4$) irradiated at ~ 650ºC and 20 dpa showed a slight (~ 20%) increase in toughness [Cli 85]. A similar value was obtained for alumina at ~ 3 dpa [Cli 85]. Tensile strength of spinel irradiated at 400–540ºC and ~20 dpa showed an increase of ~ 36% in tensile strength [Naj 90]. Thus, the degradation of the mechanical properties of the powder in the MIC will be low.

### 8.D. Decay heat

Several cases can be considered:

**1. Fresh new metals are used to fabricate a new reactor core** (that is, re-use of the activated material is not produced). The yearly-average neutron power in the μStellarator for DC 1% is 4 MW/m², about twice of typical reactors (2 MW/m² ; ~ 1 $GW_e$, and $V_p$ ~ 1000 m³). Considering iron-based material for the reactor core and a copper substrate, the decay heat at the first-wall immediately after shutdown could be estimated as ~ 0.1–1 MW/m³ , roughly twice the one in the first wall of typical fusion reactors, deduced from [Gil 15][Som 12]. In case of LOCA (loss-of-coolant accident), initially, this heat should be removed passively by the external liquid breeder. Copper produce similar decay heat than iron until the first year [Gil 15].

**2. Re-use of the activated material without isotopic separation (without reprocessing)**. Assuming that 10 recyclings are produced during 10 years, only the isotopes with half-lives of the order of 1–10 years could be considered for an initial estimation. For iron, $^{54}$Mn (~ 1 year half-life) would increase this component of the decay heat about twice, hours after shutdown, after ten years of recycling. This is estimated by the balance of accumulation and decay of $^{54}$Mn during ten years. For $^{55}$Fe (2.7 year half-life) the effect appears lower due to the low heat production in comparison to $^{54}$Mn. Thus, reuse of the material of an iron-based reactor core is considered feasible concerning decay heat, with about two-fold increase of decay heat, hours after shutdown.

For zirconium, highly active isotopes of impacting half-life (a year or few years) are not produced. Thus, a Zr reactor core could be reused many times without significant increment of decay heat hours after shutdown.

For the copper substrate, the use of the isotope $^{65}$Cu would be favourable for re-use, since $^{60}$Co (significantly accumulates; half-life ≈ 5.3 year) would not be generated.

**3. Re-use of the activated material with isotopic separation (reprocessing)**. This option seems only necessary if the reduction of the concentration of long-lived isotopes were required.

### 8.E. Neutron heating of Equi-power substrate, $\Delta T$

Regarding the neutron heating of the Equi-power substrate and the coil support near this substrate, a first approximation is obtained considering:

- The DEMO neutron heating at the first wall for ferritic-martensitic steel is 8 W/cm³ [Pal 17] for the EU DEMO reactor considered in [Pal 17].

- The specific neutron heating is similar for Fe (EUROFER) and Cu (Cu substrate, Cu in cables).

- Scaling 8 W/cm³ to the plasma surface and neutron power for a μStellarator of Case $A_{28}$ , the pulsed neutron heating is ~300 times higher in this μStellarator than in the DEMO reactor in [Pal 17] (400 MW/m² pulsed neutron power in Case $A_{28}$ and ~ 1.4 MW/m² of neutron power in such DEMO).

For this μStellarator it results in 2.4 kW/cm³ during a pulse of 0.16 s, resulting in $\Delta T_{ave}$ ~ 110 ºC at the end of the pulse. The value is high, but reasonable. The increase of temperature will be somewhat higher for the areas with peak neutron power ($\Delta T_{\rm peak}$). The value is much lower than the increase of temperature due to ionized particles, Table 2.

### 8.F. Activation

The short term and middle term (MT ≡ 100 years) activation is important for cooling during maintenance, rad-hard RH equipment, cooling during LOCA, and for requirements in the temporary hot cells.

The relative activation with respect Fe is studied next. The effect of the small (at DC 1%) different exposure time with respect Ref. [Gil 15] is neglected.

**Aluminium**: The activation of Al is 1000/10000x MT γ/decay-heat ; and ~5x LT Bq, 10000x LT γ, 1000x LT decay-heat [Gil 15], due to $^{26}$Al isotope formation, which is inconvenient. From 1–10 years, the activation of Al is much lower than iron. The activation might be acceptable for a miniature μStellarator reactor. Moreover, the reutilization/remelting of the conductor material (due to decrease of conductivity) would much reduce the volume of activated aluminium. However, the total long term activation from the long-lived $^{26}$Al isotope would not be reduced.

**Copper**: The activation of natural copper is ~10000x MT Bq/γ/decay, which could be an issue. And ~1x LT Bq/γ/decay [Gil 15] which is positive, since the long term activation of copper is equal or lower than iron. Nevertheless, **isotopic purification of copper** to remove $^{63}$Cu and use only **$^{65}$Cu** would avoid $^{60}$Co production, as deduced from [Gil 15]. A secondary reaction through the stable $^{64}$Ni would still give a small quantity of $^{63}$Ni production from high energy neutrons [Gil 15]. The elimination of $^{60}$Co and the important reduction in $^{63}$Ni would much reduce activation in the period 1–1000 yr, as deduced from [Gil 15]. Thus, copper could be considered an acceptable material concerning activation, mostly if isotopic purification and reuse is produced.

Other materials are not considered for the conductor due to lower conductivity, or high price and activation (i.e. silver).

**Monolithic coil support**:

The activation of **ODS RAFM**, i.e. Fe-14Cr-3W-0.3Ti $Y_2O_3$ will be acceptable, as widely studied for similar alloys (i.e. EUROFER). The small quantity of oxygen and yttrium is not significant for activation, as deduced from [Gil 15].

**Zirconium** has similar/lower short-term (1 year) activation than iron. γ Bq and decay-heat is 2–4 orders of magnitude lower than iron at 10 years. Activation is ~1000x MT decay-heat and ~10x MT γ/Bq ; ~100x LT Bq/γ/decay-heat [Gil 15]. The long term activation is caused by the radionuclides $^{94}$Nb (~ $10^4$ yr half-life) and $^{93}$Zr (~ $10^6$ yr half-life). **Isotopic purification of Zr** to use only $^{90}$Zr (the most abundant natural isotope) would almost eliminate the production of $^{94}$Nb and $^{93}$Zr, as deduced from [Gil 15]. Thus, zirconium could be acceptable for this reactor concept, mainly if isotopic purification and material reuse is produced. The other alloying materials of the particular ODS 49Zr-30Fe-10Cr-5Cu-5Ti (Fe, Cr, Ti) are of low activation, except for Cu, though $^{65}$Cu could be used, or Cu changed in the alloy.

### 8.G. Discussion on neutronics, difficulties

The need to replace the solid materials directly exposed to fusion neutrons at periods approximately proportional to the plasma surface/size of the device (and inversely to the fusion power) can be deduced from the reactor studies of ARIES-CS (replacement each 2.8 yr), HSR22 (each 8 yr) and FFHR (10 yr) [Lyo 07][Ama 99][Sag 05]. For the μStellarator, the replacement of the monolithic coil support is required about each year for 1% duty cycle. The higher complexity and accuracy of the tiny reactor core in comparison to the large simpler blankets is the main difference.

The small size and the numerous (half)periods facilitate their serial production/refurbishment, i.e. by casting. Actually, casting of a coil support was demonstrated [Que 23]. Also, re-use of the activated materials could be considered, as proposed in [Pac 12][Zuc 09][Mas 07]. Certainly, this process would have to be produced under special radiation-capable smelting/3D-printing facilities, somewhat similar to reprocessing facilities for fission.

The increase of resistivity of aluminium with radiation require further studies, in particular to discover if the increment of resistivity saturates, alike copper. This matter appears advantageous for copper.

Neutron damage resulted ~ 0.1 dpa/day on the conductors for duty cycle 1%. Thus, in-site (inside the monolithic coil support) re-conditioning or re-melting of the conductor (Cu or Al) under vacuum would be convenient. This method of coil production is used in the industry for the production of certain contorted coils for radiofrequency tempering.

The reduction of duty cycle, i.e. to 0.1%, would increase lifetime of all the elements 10-fold. This would be important for the initial development phases of the concept, but, it is not reasonable for the competitive production of heat or electricity.

Electrical degradation and RIC of the conductor insulation appear to be negligible for this application. The conductor will be insulated by MIC. MIC keeps sufficient electric resistance under neutron radiation. Fatigue behaviour of the MIC insulation need further studies and tests.

Wastes due to activation of the coil support and Cu conductors appears acceptable.

## 9. Discussion. Challenges.

In this section, the **further studies and tests required**, the **difficulties and challenges** of the concept and the current **uncertainties** are described and summarized.

Next, a matter is discussed in each paragraph.

The most important concern on the concept is the lack of experimental demonstration of various of the key elements involved in the μStellarator concept. Presently, it implies low Technology Readiness Level of the concept.

Pulsed regime has negative effects, particularly on thermo-mechanical fatigue and energy losses during ramp-up. Nonetheless, it appears hardly feasible a simple and tiny commercial (net energy production) magnetic fusion reactor if not pulsed, due to thermal/practical limits of the cooling systems.

The Distributed Divertor concept has only been validated theoretically in an initial manner [Que 25]. Advanced calculations including molecular interactions, possible sputtering and plasma purity would be relevant. Experimental validation would be also very valuable, since excessive impurity influx, excessive non-uniformity of the heat power on the surface, or difficult positioning of the Equi-power surface with respect the plasma would challenge the Distributed Divertor concept. Nevertheless, also, island divertors under full detachment, high edge radiation and divertor-leg-sweeping might distribute the ionized particle power on the full surface of the vacuum vessel. This is a back-up option for the divertorial power extraction in μStellarators.

Helium pumping is not considered an issue due to the short pulse length (5 $\tau_E$) which gives ≈ 8% burning fraction for the Case $A_{28}$, together with the potential (partial) pumping of He by lithium.

The non-conformal shape of the Distributed Divertor [Que 25] with respect the plasma implies an increase of length of the conductors with respect a circular shape. It represents a slight increase of resistivity of the coils. This increase has not been included in the estimations.

ODS Fe-14Cr-3W-0.3Ti ($Y_2O_3$) and ODS Zr Alloy $Y_2O_3$ (i.e. ODS 49Zr-30Fe-10Cr-5Cu-5Ti-1$Y_2O_3$) are capable for Case $A_{28}$ (and Case $A'_{20}$) including the effect of mechanical fatigue. Only maraging steel and perhaps the ODS Zr Alloy $Y_2O_3$ could cope with the stresses in Case $A_{20}$. However, maraging steel suffers from large swelling and high activation, Table 3.

A fatigue ratio coefficient has been included in the mechanical estimations (FR = 0.7, Section 7.E), considering non-reversal fatigue, proper surface treatment and high quality design. Nevertheless, further refinement and experiments are needed.

Large thermal gradients in the reactor core will contribute to mechanical stresses and thermo-mechanical fatigue in the structure, due to the pulsed regime.

Neutron-lifetime of the coil support structure would be about one year for duty cycle 1% (for the Case $A_{28}$). Thus, the full structure should be re-fabricated each year, preferably by reusing the same activated material in a casting o 3D printing system.

Numerous conductors in the inter-cable matrix/structure is the reference option, Fig. 11, to decrease the stresses in the insulation and distribute the forces in the inter-cable matrix/structure.

The mechanical stress and fatigue of the insulation has not been calculated or simulated by FEA.

The radiation resistance of the electrical insulation under μStellarator conditions has been preliminarily assessed and seems feasible, Section 8.C.

Splitting the reactor in periods, half-periods [Que 18] or further divisions is favourable for: **a)** Maintenance, i.e. replacement of potential failed (half)periods, and, **b)** Lower the production cost by serial production of half-periods. However, splitting complicates the attachment between the splitted sectors.

Only ECRH heating appears appropriate for this minute device (lack of space). Undoubtedly, special concentration of ECRH beams will be required to fulfil the thermal resistance of windows (Fig. 1), as already defined in Ref. [Que 10] and resembling to the NIF (National Ignition Facility) windows. Gyrotrons of the required frequency for maximum cut-off density are still experimental and have low efficiency, Section 4.B.

For 1% DC, cooling of the Equi-power substrate is feasible and relatively simple for Case $A_{28}$. For 5% DC and Case $A_{20}$ the cooling rate required on the whole interior of the vacuum vessel is similar to the one in ITER divertor targets – much more demanding.

The existence of plasmas of high beta (>=20%) in large aspect ratio stellarators has only been proved theoretically, i.e. [Bov 08]. Also, a magnetic configuration of $\beta_{\text{lim}}$ = 14% and $A$ = 20 has been found by the authors as a preliminary configuration for μStellarators, Section 3.B. $\beta_{\text{lim}}(\%) > A$ would be advantageous.

The central plasma pressure in $A_{28}$ μStellarator would be ~ 45 MPa, which is much higher than in Alcator C-Mod [Hug 18]. The much higher magnetic field and higher beta in μStellarators may allow such high pressure gradients.

Fast particle behaviour and resistive MHD has not been studied.

Obtaining a magnetic configuration integrated with the Distributed Divertor (or island divertor) require future studies. The Distributed Divertor was theoretically initially demonstrated for a quasi-helical magnetic configuration (HSX, [Que 25]), not a QI configuration. Indeed, QI configurations may be preferred due to simpler plasma positioning and reduced Shafranov shift, which are of particular importance for short-pulse devices.

Pure natural lithium gave promising result as breeding material (TBR ≈ 1) for μStellarators in our initial MCNP estimations, Section 8.A. Sufficient TBR should be confirmed in a more realistic toroidal geometry.

Lithium safety due to the relatively large volume of lithium in the pool (~ 700 $m^3$ in the scheme in Fig. 1, but, it could be much reduced by optimization) is a concern. Li-Pb and solid pebbles have not been studied.

The tritium cycle and anti-permeation measures have not been evaluated. They should be similar to the common methods for standard reactors.

Remote maintenance procedures have not been defined. The (remote) maintenance procedures for stellarators and other devices described in Refs. [Que 24b][Que 11][Que 15][Que 18], which are aimed for HELIAS reactors, IFMIF, EU-DEMO, UST_2 stellarator and i-ASTER stellarator reactor, will be applied to μStellarators. The small size of the μStellarator periods and the potential for in-site (in the reactor vessel) remelting/refurbishement of the neutron-damaged reactor core will much simplify and give reasonable cost of the maintenance procedures.

The cost of the monolithic coil support plus coils-layer (they support the Lorentz forces) is assumed scaling approximately proportional to the magnetic field $B$ and to the square root of the magnetic volume, following [Gre 08]. Nevertheless, a recent study for tokamak reactors states a cost scaling with the square of $B$ [Fed 24]. The exact intermediate point of cost scaling should be better evaluated.

### 9.A. Prospects of heat and electricity production

The prospect of net energy production in μStellarators can be interpreted as: **a)** Net energy/heat production, that is, $\gamma \approx 1$ (see Section 2.B), **b)** Reasonable electricity production with i.e. $\gamma \approx 2$, as contemplated in this study.

Thus, from this initial study, reasonable electricity production (interpretation 'b') appears feasible if $A \sim \beta_{\text{lim}}\%$ at high $\beta_{\text{lim}}$ and the other thermo-mechanical parameters in the study are achieved.

Even if significant reasonable electricity production were not achieved (interpretation 'b') due to any unknown, still, huge net heat production (interpretation 'a') could be obtained in μStellarators. Heat would be intended for industrial or urban heating.

### 9.B. Duty cycle and competitiveness of the reactor

Not only net energy production is important for a commercial fusion reactor. Also, production cost equal or lower than other energy sources is essential.

For duty cycle 1%, simple substrate-cooling pipes and low fluid speed is required, see Section 5.C. At 1% DC, the electric power produced by the reactor would be ~ 30 $MW_e$ for the Case $A_{28}$ and ~ 60 $MW_e$ for the Case $A_{20}$, which is considered sufficient for certain niches of market.

For duty cycle 5%, the electric power produced by the reactor would be ~ 150 $MW_e$ for the Case $A_{28}$ and ~ 300 $MW_e$ for the Case $A_{20}$. Certainly, for the Case $A_{20}$, the full internal surface of the reactor (the Equi-power substrate) has to be cooled at relatively high speed and use swirling-pipes, Section 5.C. These design and duty cycle would be competitive in wider markets.

The flowing liquid metal alternatives mentioned in Section 5 might allow higher power exhaust than solids.

## 10. Advantages of the concept

A µStellarator is tiny, simple and composed of relatively common materials. Thus, minimizing the cost during the development phase and during construction. This is much effectively achieved in a very small ($V_p$ < ~ 10 $m^3$) commercial fusion reactor without the complexity of superconducting coils, cryostat, cryo-shieldings, several nested layers of elements (blankets, vacuum vessel, coils, cryostat), large remote maintenance equipment, large reactor building and large hot cells.

Moreover, a miniature reactor allows frequent magnetic configuration improvements, favoured by the small size and the frequent need of replacement due to neutron damage.

Though large (>200–400 $m^3$ plasma volume) steady-state high-power (> 3–5 $GW_e$) superconducting fusion reactors may be required to competitively supply the future social energy needs, still, µStellarator reactors may keep certain niches of market.

The production of certain profit in certain niches of market during the upswing phase is essential for the straightforward scale-up of a new product, as happened with photovoltaic solar or fission energy. Indeed, commercial profit is envisaged from the µStellarator concept by the production of industrial heat, tritium/$^3$He, and likely, electricity production. Breeding $^{233}$U for thorium fission breeder reactors (i.e. in India) at low Q and relaxed parameters is also envisaged as a market niche. This prospect of profit from the initial development stages should be stimulating for companies.

Additionally, advancement on the knowledge of plasma physics and materials would be also produced in µStellarators.

## 11. Summary and conclusions

A new concept of ultra-high-field miniature commercial stellarator reactor has been formulated and developed, called µStellarator (microStellarator), able to produce industrial and domestic heat and electricity, and thus, it is a commercial reactor. Additionally, it would speed up the innovation cycle and advance plasma physics. It works in pulsed regime, use resistive magnets and the breeding material is located outside the resistive magnets (*transposed* stellarator reactor). It has some similarities, but also many differences, to the old Riggatron tokamak concept [Ros 84]. Another important feature is the use of the innovative non-resonant Distributed Divertor. The Distributed Divertor uniformly distributes the ionized-particle power on the full internal vacuum vessel (Equi-power surface) [Que 25]. High beta limit (10%–40%) stellarators, large plasma aspect ratio and other conditions have been considered in the study, to be able to achieve reasonable net electricity production (~ 50% recirculated electricity to the coils). A lithium film as CPS or boron coating is planned to pump ionized particles on the Equi-power surface. Liquid Li PFCs will favour the plasma purity, improve confinement due to low recycling regime and increase the lifetime of the solid substrate.

The resistive magnets are massive and the coils have variable cross-section to decrease the Joule-effect losses in the coils to achieve reasonable net electricity production.

The working parameters of these ultra-high-field commercial reactors are studied. Plots on the relevant physics and engineering parameters are generated for a wide range of plasma volumes, being the relevant interval 2–4 $m^3$. Cases have been found (i.e. $A_{28}$, $A'_{20}$ and perhaps $A_{20}$) with potential feasible parameters under the condition of reasonable net electricity production.

The pulsed (~0.1–0.2 s) divertorial heat power for Cases $A_{40}$ to $A_{20}$ ranges from ~ 60 to 190 $MW/m^2$ respectively. This pulsed heat is admissible for copper and for tungsten Equi-power substrate, for certain Cases and conditions. Low recycling lithium on the substrate appears feasible only for the Cases $A_{28}$ and $A'_{28}$ using longitudinal ridges, and perhaps $A_{28}$ without ridges. Also, the heat level is admissible for the Case $A_{40}$, but, this Case seems hardly possible concerning plasma physics. Flowing liquid lithium may avoid the use of ridges, but, with higher complexity in other aspects.

Table 4 summarises the working parameters of two of the most promising Cases.

| Parameter | Case $A_{28}$ | Case $A_{20}$ |
|---|---|---|
| $V$ | 3 $m^3$ | 2 $m^3$ |
| $B$ | 13.5 T | 18.8 T |
| $R$ | 4.9 m | 3.4 m |
| $a$ | 0.17 m | 0.17 m |
| $A$ | 28 | 20 |
| Plasma surface | 35 $m^2$ | 24 $m^2$ |
| $n_{line}$ | $1.2 \times 10^{22}$ $m^{-3}$ | $1.7 \times 10^{22}$ $m^{-3}$ |
| $T_0$ | 13.7 keV | 13.7 keV |
| Fusion energy gain Q | $Q \rightarrow \infty$ (ignition) | $Q \rightarrow \infty$ (ignition) |
| Fusion power (pulsed) | 18.1 GW | 22.6 GW |

| Electricity production, DC 1% | ~ 30 $MW_e$ | ~ 60 $MW_e$ |
|---|---|---|
| Electricity production, DC 5% | ~ 150 $MW_e$ | ~ 300 $MW_e$ |
| $h_E$ (ISS04) | 1.4 | 1 |
| $<\beta>$ | 28 % | 20 % |
| $\tau_E$ | 0.033 s | 0.024 s |
| Pulse length $t_p$ | 0.16 s | 0.12 s |
| Load on Equi-power surf. (pulsed, no ridges) | ~100 MW/m² | ~190 MW/m² |
| Neutron wall load (pulsed, average) | ~400 MW/m² | ~760 MW/m² |
| $\delta = (\varepsilon + \psi)\ a$, shell thickness | 0.255–0.3 m ; (1 + 0.5–0.75)$a$ | |
| $P_{\mathrm{coils}}$ (pulsed) | 3.2 GW | 4.2 GW |
| Rough max. stress in coil support | ~ 480 MPa | ~ 900 MPa |
| Ref. materials of reactor core | Copper conductor + ODS steel or ODS Zr alloys | |

**Table 4** Parameters of two of the most promising Cases, being the Case $A_{20}$ more engineering challenging and Case $A_{28}$ more physics demanding.

No unsurmountable difficulties have been found for this concept of ultra-high-field miniature commercial stellarator reactor, but, certainly, further studies and experiments are needed to better evaluate the concept. The current main concerns are: the feasibility of high beta limit equal to the plasma aspect ratio at large aspect ratios, the practical feasibility of the Distributed Divertor or a high-power island divertor, the strength of the inter-cable matrix, and the endurance of the electrical insulation under mechanical fatigue. These matters, together with enhanced neutronics calculations, estimations on the effects and power consumption/storage during ramp-up/down and better integration of the elements, remain for future works.

## Acknowledgements

The work was funded by the 'Agencia Estatal de Investigación' (AEI), the 'Ministry of Science, Innovation and Universities', and by the 'Fondo Europeo de Desarrollo Regional' European Union (FEDER EU), under the Grant n. PID2021-123616OB-I00, for the project "Study of improved stellarator assemblies consistent with proper in-vessel components for viable high-field stellarator reactors", and by the grant PID2022-137869OB-I00. Calculations were done in Uranus, a supercomputer cluster located at Universidad Carlos III de Madrid and funded jointly by EU-FEDER and the Spanish Government via Project No. UNC313-4E-2361, ENE2009- 12213-C03-03, ENE2012-33219 and ENE2015-68265.

The authors are grateful to J. Nührenberg and C. Nührenberg (Max Planck Institute for Plasma Physics) for supplying the QIP6 magnetic configuration [Sub 06] years ago, and allowing its utilization as optimization 'seed' for this publication; and to Raul Sanchez (UC3M ; MHD, VMEC, COBRA). And to CIEMAT researchers Daniel Alegre (for advice on divertorial matters, materials), Edilberto Sanchez (parallel computing), and Emilio Blanco (electromagnetism, pulsed fields).

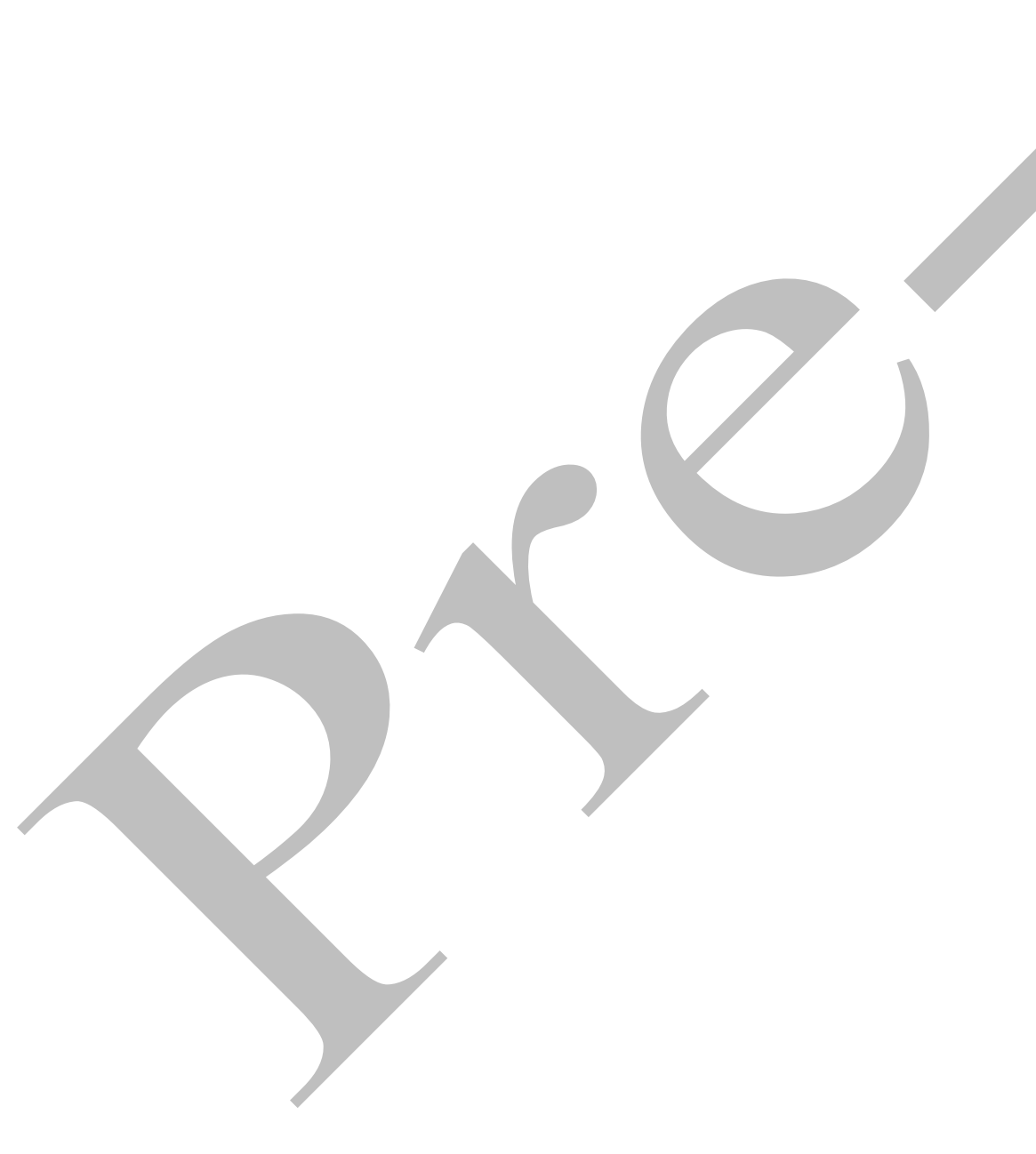